\documentclass[journal,onecolumn]{IEEEtran}

\ifCLASSINFOpdf
   \usepackage[pdftex]{graphicx}
\else
\fi
\usepackage[cmex10]{amsmath}
\usepackage{array}
\usepackage[cmex10]{amsmath}
\usepackage{xfrac}
\usepackage{array}
\usepackage{mdwmath}
\usepackage{amsfonts}
\usepackage{amssymb}
\usepackage{amsbsy}
\usepackage{amsthm}
\usepackage{tikz}
\usetikzlibrary{arrows}
\newtheorem{theorem}{Theorem}
\usepackage{setspace}
\begin{document}
%
\title{Capacity Bounds and a Certain Capacity Region for Special Three-Receiver Broadcast Channels with Side Information}
%
%
%
 
\author{Sajjad~Bahrami,~\IEEEmembership{Student Member,~IEEE,} and
        Ghosheh~Abed~Hodtani,~
\thanks{Sajjad Bahrami is with the Department
of Electrical Engineering, Ferdowsi University of Mashhad, Mashhad, Iran (e-mail: sajjad.bahrami.1990@gmail.com).}
\thanks{Ghosheh Abed Hodtani is with the Department of Electrical Engineering, Ferdowsi University of Mashhad, Mashhad, Iran (e-mail: ghodtani@gmail.com).}
}

\maketitle

\begin{abstract}
The fact that the results for $2$-receiver broadcast channels (BCs) are not generalized to the $3$-receiver ones is of information theoretical importance. In this paper we study two classes of discrete memoryless BCs with non-causal side information (SI), i.e. multilevel BC (MBC) and 3-receiver less noisy BC. First, we obtain an achievable rate region and a capacity outer bound for the MBC. Second, we prove a special capacity region for the $3$-receiver less noisy BC. Third, the obtained special capacity region for the $3$-receiver less noisy BC is extended to continuous alphabet fading Gaussian version. It is worth mentioning that the previous works are special cases of our works.
\end{abstract}


%
\IEEEpeerreviewmaketitle

\section{Introduction}
%
%
%
%
\IEEEPARstart{B}{roadcast} channel is an important multiuser channel, first introduced by Cover \cite{cover1972}. Bergmans obtained an achievable rate region for degraded $2$-receiver BC \cite{bergmans} using superposition coding and Gallager \cite{gallager} and Ahlswede-K\"{o}rner \cite{ahlswede&korner} showed that it is optimal. Special classes of BCs have been studied in \cite{elgamal} and \cite{Korner&Marton1977}. By now, the best known capacity bounds for general $2$-receiver BC are given by Marton \cite{martoninnerbound} and Nair-El Gamal \cite{nayerelgamouter}. The results for BCs with two receivers are not generalizable to the BCs with more than two receivers
and finding a closed form result for a broadcast channel with arbitrary number of receivers is not a straightforward problem. So some works focused on some special classes of BCs with more than two receivers such as multilevel broadcast channel \cite{nairelg2009}  and 3-receiver less noisy broadcast channel \cite{nairwang2011}.
For the first time, Borade et al. studied BCs with more than two receivers \cite{borade}. They guessed the straightforward extension of capacity region for $2$-receiver BC with degraded message sets obtained by K\"{o}rner and Marton \cite{Korner&Marton1977}, to multilevel broadcast channel, however Nair and El Gamal \cite{nairelg2009} showed that it is not in general optimal. Later, Nair and Wang \cite{nairwang2011} obtained the capacity region of the 3-receiver less noisy BC. 

Shannon studied channels with side information (SI) \cite{shanon}, and found the capacity region of the single user channel when causal SI is available at the transmitter. Single user channel, when non-causal SI is available only at the transmitter, was studied by Gel'fand-Pinsker \cite{gelfandpinsker}. The results of \cite{gelfandpinsker} was generalized by Cover and Chiang \cite{coverchiang} to the case where non-causal SI is available at both transmitter and receiver. Also single user channel with partial channel state information at the transmitter (CSIT) was studied in \cite{Jadid2}. Multiuser channels e.g. multiple access channels, BCs, interference channels and relay channels in presence of SI were studied in \cite{JADID1}, \cite{steinberg2005}, \cite{Monemizadeh} and \cite{Osmani}, respectively and some works on multiuser channels with partial SI  can be found in \cite{bahmani1}, \cite{bahmani2}, \cite{bahmaniISIT2012} and \cite{Jadid3}.

BC with SI was investigated in \cite{steinberg2005}, in which inner and outer capacity bounds and capacity region for different cases of SI presence in degraded BC were obtained. The authors in \cite{Steinberg&Shamai} investigated general BC in presence of non-causal SI at the transmitter, actually they considered SI in \cite{martoninnerbound}. Later the authors in \cite{Marvasti} considered SI non-causally available at the transmitter in multiuser channels and obtained  their capacity bounds, specifically for general BC they considered SI for both \cite{martoninnerbound} and \cite{nayerelgamouter} and concluded the result of \cite{Steinberg&Shamai}.

Costa \cite{Costa} studied Gaussian channels with SI, actually he investigated Gaussian version of \cite{gelfandpinsker}. Also a more general version of Costa's theorem was studied in \cite{anzabinejad}. The authors in \cite{Kim} extended the results in \cite{Costa} to multiuser channels. Two-way channel, first studied by Shannon \cite{**Shannon}, was investigated in \cite{LetterFarsani} by extending \cite{Costa}. As we know, in addition to noise, fading is another problem in wireless communications and some works have studied this phenomenon in \cite{Goldsmith}, \cite{Tse} and \cite{farsani}. One of the most efficient works on capacity bounds for fading BC with partial CSIT is mentioned in \cite{farsani}. As it is described in \cite{farsani}, this approach has some advantages such as controlling the amount of CSIT, no need for necessarily separating the channel into parallel sub-channels to analyze it, which is not applicable for some channels such as interference channels, and also it provides a simple usage of entropy power inequality (EPI), which is brought in \cite{dembo}, for outer bound proof.

\textbf{Our works:} 
\begin{itemize}
\item First, an achievable rate region is obtained for discrete memoryless MBC with SI non-causally available at the transmitter: 

It can be easily seen that the obtained achievable rate region is reduced to that of \cite{nairelg2009} when we consider the distribution as in \cite{nairelg2009} and have no SI, and subsumes Steinberg's achievable rate region for $2$-receiver degraded BC with SI as its special case. 

\item Second, considering SI available at both the transmitter and receivers, we obtain a  capacity outer bound for MBC.

\item Third, an achievable rate region is derived for discrete memoryless $3$-receiver less noisy BC with SI non-causally available at the transmitter:  

For this channel, we do as for MBC and again our achievable rate region reduces to that of \cite{nairwang2011} when we consider the distribution as in \cite{nairwang2011} and there is no SI.

\item Fourth, we obtain the capacity region for discrete memoryless $3$-receiver less noisy BC with non-causal SI available at the transmitter and receivers.

\item Fifth, capacity bounds for the fading Gaussian $3$-receiver less noisy BC with partial CSIT are obtained:

Although all works mentioned above are discrete alphabet cases, we use both the capacity region found in our paper for discrete memoryless $3$-receiver less noisy BC and an efficient scheme in order to extend this region to channel with discrete time and continuous alphabet fading Gaussian channel, as stated in \cite{gallagerbook}, to obtain capacity bounds for the fading Gaussian $3$-receiver less noisy BC.
\end{itemize}
 \textbf{Notation:} We show random variables and their realizations by uppercase and lowercase letters, respectively, e.g. $x$ is a realization of $X$, also the $n$-sequence of a random variable is illustrated by $X^n$ and its realization is denoted by $x^n$. Let $X_i^n$ be the sequence $\left(X_i, X_{i+1},\cdots, X_n\right)$ and $\mathcal{S}$, $\mathcal{X}, \mathcal{Y}_1, \mathcal{Y}_2$ and $\mathcal{Y}_3$ be finite sets which show alphabets of random variables. Furthermore, throughout the paper $\bar{\alpha}\left( .\right)$, $\bar{\beta}\left( .\right)$, $\bar{\vartheta} \left( .\right)$ and $\bar{\gamma}\left( .\right)$ mean $1-\alpha \left( .\right)$, $1-\beta \left( .\right)$, $1-\vartheta \left( .\right)$ and $1-\gamma \left( .\right)$, respectively, also $E\{ . \}$ denotes expectation operator and we use $\psi \left( x \right)=\log \left(1+x\right)$.
 
\textbf{Paper Organization:} The remainder of this paper is organized as follows. Section II provides some preliminaries and definitions. In section III, we study MBC with non-causal SI and section IV is devoted to $3$-receiver less noisy BC with non-causal SI. In both sections II and III discrete alphabet channels are studied. 
In section V, we investigate fading Gaussian 3-receiver less noisy broadcast channel and in section VI, we conclude the paper.

 




\section{Basic Definitions and Preliminaries}
In this section, two lemmas and some definitions are reviewed.
\subsection{Preliminaries} 
 Note that in this paper we use the weak notion of typicality.  

\textbf{Covering Lemma}: Consider $\left(U,S,V\right)\sim p\left(u,s,v\right)$ and $\epsilon' < \epsilon$. Also $\left(U^n,S^n\right)\sim p\left(u^n,s^n\right)$ is a pair of random sequences with the property:
{\small
\begin{equation}
lim_{n\rightarrow\infty} p\{\left(U^n,S^n\right)\in \tau_{\epsilon^{'}}^{(n)} \left(U,S\right)\}=1, \nonumber
\end{equation}}   
and assume $V^n\left(m\right), m\in \mathcal{A}$, where $2^{nR}\leq\mid \mathcal{A}\mid$, are random sequences, conditionally independent of each other and of $S^n$ given $U^n$, each distributed according to $\sideset{}{_{i=1}^n}\prod p_{V\mid U}\left(v_i\mid u_i\right)$. Now, there exists $\delta\left(\epsilon\right)$ that tends to zero as $\epsilon \rightarrow 0$ such that:
{\small
\begin{equation}
lim_{n\rightarrow\infty} p\{\left(U^n, S^n, V^n\left(m\right)\right)\notin \tau_\epsilon^{(n)} for~ all~ m\in \mathcal{A}\}=0, \nonumber
\end{equation}}          
if we have:
{\small
\begin{equation}
R>I\left(V;S\mid U\right) + \delta\left(\epsilon\right). \nonumber
\end{equation}}
The proof is brought in \cite{elgama2011}. This lemma is used for encoding error analysis.

\textbf{Packing Lemma}: Consider $\left(U,X,Y\right)\sim p\left(u,x,y\right)$. Also assume  $\left(\tilde{U}^n,\tilde{Y}^n\right)\sim p\left(\tilde{u}^n,\tilde{y}^n\right)$ is a pair of arbitrary distributed random sequences. Let $X^n\left(m\right)$, $m\in\mathcal{A}$ and $\mid \mathcal{A}\mid\leq2^{nR}$ be random sequences, each distributed according to $\sideset{}{_{i=1}^n}\prod p_{X\mid U}\left(x_i\mid \tilde{u}_i\right)$. Now consider $X^n\left(m\right), m\in \mathcal{A}$, is arbitrary dependent on other $X^n\left(m\right)$ sequences and is pairwise conditionally independent of $\tilde{Y}^n$ given $\tilde{U}^n$. Now, there exists $\delta\left(\epsilon\right)$ that tends to zero as $\epsilon \rightarrow 0$ such that:
{\small
\begin{equation}
lim_{n\rightarrow\infty} p\{\left(\tilde{U}^n, X^n\left(m\right), \tilde{Y}^n\right)\in \tau_\epsilon^{(n)} for~ some~ m\in \mathcal{A}\}=0, \nonumber
\end{equation}}
if we have:
{\small
\begin{equation}
R<I\left(X;Y\mid U\right) - \delta\left(\epsilon\right). \nonumber
\end{equation}}
The proof can be found in \cite{elgama2011}. This lemma is used for decoding error analysis.
\subsection{Definitions}
In this subsection some basic definitions and models of the channels are presented.

\textbf{Definition 1}: In a broadcast channel in presence of SI, $p\left(y,z\mid x,s \right)$, the channel from $X$ to $Z$ is said to be a degraded version of the channel from $X$ to $Y$ if we have the following Markov chain conditioned on every $s\in\mathcal{S}$:
{\small
\begin{equation}
X,S \rightarrow Y \rightarrow Z. \nonumber
\end{equation}}

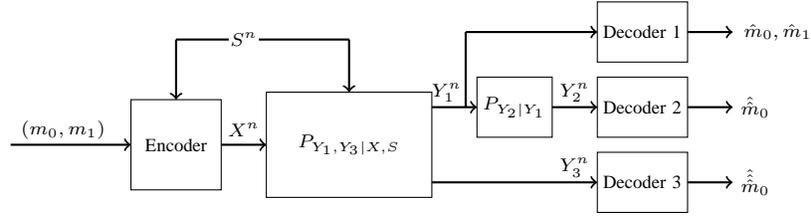
\begin{figure}
\centering
\begin{tikzpicture}
\draw [fill=white] (0,0) rectangle (1.2,1.2);
\node at (0.6,0.6) {\scriptsize Encoder};
\draw [->, thick] (-1.6,0.6)--(0,0.6);
\node at (-0.9,0.8) {\scriptsize $\left(m_0, m_1\right)$};
\draw [->, thick] (1.2,0.6)--(1.8,0.6);
\node at (1.5,0.8) {\scriptsize $X^n$};
\draw [fill=white] (1.8,-0.1) rectangle (4,1.3);
\node at (2.9,0.6) {\scriptsize $P_{Y_1, Y_3\mid X, S}$};
\draw[->,thick] (.6,2)--(.6,1.2);
\draw[->,thick] (2.9,2)--(2.9,1.3);
\node at (1.5,2) {\scriptsize $S^n$};
\draw[thick] (1.3,2)--(.6,2);
\draw[thick] (1.7,2)--(2.9,2);
\draw[->, thick] (4,1.1)--(4.6,1.1);
\draw[->, thick] (4,.1)--(6.2,.1);
\draw[fill=white] (4.6,.7) rectangle (5.6,1.5);
\node at (5.1,1.1) {\scriptsize $P_{Y_2\mid Y_1}$};
\draw[->, thick] (5.6,1.1)--(6.2,1.1);
\node at (4.2,1.3) {\scriptsize $Y_1^n$};
\node at (5.9,1.3) {\scriptsize $Y_2^n$};
\node at (5.9,.3) {\scriptsize $Y_3^n$};
\draw [thick] (4.45, 1.1)--(4.45,2.1);
\draw [->,thick] (4.45, 2.1)--(6.2,2.1);
\draw (6.2,1.7)rectangle(7.4,2.5);
\draw (6.2, .7)rectangle(7.4,1.5);
\draw (6.2, -.3)rectangle(7.4,.5);
\node at (6.8,2.1) {\scriptsize Decoder 1};
\node at (6.8,1.1) {\scriptsize Decoder 2};
\node at (6.8,.1) {\scriptsize Decoder 3};
\draw[->,thick] (7.4,2.1)--(8,2.1);
\draw[->,thick] (7.4,1.1)--(8,1.1);
\draw[->,thick] (7.4,.1)--(8,.1);
\node at (8.6, 2.1) {\scriptsize $\hat{m}_0, \hat{m}_1$};
\node at (8.3, 1.1) {\scriptsize $\hat{\hat{m}}_0$};
\node at (8.3, .1) {\scriptsize $\hat{\hat{\hat{m}}}_0$};
\end{tikzpicture}
\caption{Multilevel broadcast channel with side information.}
\label{Fig1}
\end{figure}

\textbf{Definition 2}:
Multilevel broadcast channel with SI, which is shown by $\left(\mathcal{X, S}, \mathcal{Y}_1, \mathcal{Y}_2, \mathcal{Y}_3 , p\left(y_1, y_3\mid x, s\right), p \left(y_2\mid y_1\right) \right)$, 
is a $3$-receiver BC with two degraded message sets in which the relation between channel input $X$, side information $S$, and channel outputs $Y_1$ and $Y_3$ is shown by $p\left(y_1, y_3\mid x, s\right)$ while $p\left(y_2\mid y_1\right)$ shows the output $Y_2$ as the degraded version of $Y_1$. Also, the random variable $S$ shows SI and is distributed over the set $\mathcal{S}$ according to $p_S\left(s\right)$. Let messages $m_0\in\mathcal{M}_0$ and $m_1\in\mathcal{M}_1$ be in two independent message sets, where $m_0$ is the common message and is sent to all the receivers and the private message $m_1$ is only sent to $Y_1$. Channel model is depicted in Fig. \ref{Fig1}.

A $\left(n, 2^{nR_0}, 2^{nR_1},\epsilon\right)$ $2$-degraded message set code for the MBC with SI, consists of an encoder map as below:
{\small
\begin{equation}
e: \{1,2,\cdots ,M_0\}\times \{1,2,\cdots ,M_1\} \times \mathcal{S}^n \rightarrow \mathcal{X}^n, \nonumber
\end{equation}}
and decoding maps as below:
{\small
\begin{eqnarray}
d_{y1}: \mathcal{Y}_1^n \rightarrow \{1,2,\cdots ,M_0 \} \times \{1,2,\cdots ,M_1\} \nonumber\\
d_{y2}: \mathcal{Y}_2^n \rightarrow \{1,2,\cdots ,M_0\} \nonumber\\
d_{y3}: \mathcal{Y}_3^n \rightarrow \{1,2,\cdots ,M_0\}, \nonumber
\end{eqnarray}}
such that $P_e^{(n)}\leq \epsilon$ , i.e.:

{\small
\begin{equation}
\frac{1}{M_0M_1}\sum_{m_0=1}^{M_0}\sum_{m_1=1}^{M_1} \sum_{s^n\in \mathcal{S}^n} p\left(s^n\right) p \left\{ d_{y1}\left(y_1^n\right)\neq \left(m_0, m_1\right) ~or~ d_{y2}\left(y_2^n\right)\neq m_0 ~or~ d_{y3}\left(y_3^n\right)\neq m_0 \mid s^n,x^n \left( m_0,m_1,s^n\right)\right\}\leq \epsilon. \nonumber
\end{equation}}
The code's rate pairs are defined as:
{\small
\begin{equation}
\left(R_0, R_1\right)=\frac{1}{n}\left(\log M_0, \log M_1\right). \nonumber
\end{equation}}

If for any $\xi>0$ there exists an integer $n_0$ such that for all $n\geq n_0$ there is a $\left( n,2^{n\left( R_0-\xi\right)},2^{n\left( R_1-\xi\right)},\epsilon \right)$ code for $\left( p\left( y_1,y_3\mid x,s\right),p\left( y_2\mid y_1\right) \right)$, then a rate pair $\left( R_0,R_1\right)$ is $\epsilon$-achievable. The capacity region is the union of the closure of all $\epsilon$-achievable rate pairs.

\textbf{Definition 3}:
In a broadcast channel in the presence of SI, $p\left( y,z\mid x,s\right)$, the channel from $X$ to $Y$ is said to be less noisy than the channel from $X$ to $Z$ if we have:
{\small
\begin{equation}
I\left(U;Y\mid S=s\right)\geq I\left(U;Z\mid S=s\right) ~~ ;\forall s\in \mathcal{S} \nonumber \\
\mathrm{and} ~\forall p\left(u,x,y,z\mid s\right)=p\left(u\mid s\right)p\left(x\mid u,s\right) p\left(y,z\mid x,s\right). \nonumber 
\end{equation}}

\begin{figure}
\centering
\begin{tikzpicture}
\draw [fill=white] (0,0) rectangle (1.2,1.2);
\node at (0.6,0.6) {\scriptsize Encoder};
\draw [->, thick] (-1.6,0.6)--(0,0.6);
\node at (-0.9,0.8) {\scriptsize $\left(m_1, m_2, m_3\right)$};
\draw [->, thick] (1.2,0.6)--(1.8,0.6);
\node at (1.5,0.8) {\scriptsize $X^n$};
\draw [fill=white] (1.8,-0.5) rectangle (4,1.7);
\node at (2.9,0.6) {\scriptsize $P_{Y_1, Y_2, Y_3\mid X, S}$};
\draw[->,thick] (.6,2)--(.6,1.2);
\draw[->,thick] (2.9,2)--(2.9,1.7);
\node at (1.5,2) {\scriptsize $S^n$};
\draw[thick] (1.3,2)--(.6,2);
\draw[thick] (1.7,2)--(2.9,2);
\draw[->, thick] (4,1.5)--(4.7,1.5);
\draw[->, thick] (4,-.3)--(4.7,-.3);
\draw[->, thick] (4,.6)--(4.7,0.6);
\node at (4.4,1.7) {\scriptsize $Y_1^n$};
\node at (4.4,.8) {\scriptsize $Y_2^n$};
\node at (4.4,-.1) {\scriptsize $Y_3^n$};
\draw (4.7,1.1)rectangle(5.9,1.9);
\draw (4.7, .2)rectangle(5.9,1);
\draw (4.7, -.7)rectangle(5.9,0.1);
\node at (5.3,1.5) {\scriptsize Decoder 1};
\node at (5.3,.6) {\scriptsize Decoder 2};
\node at (5.3,-.3) {\scriptsize Decoder 3};
\draw[->,thick] (5.9,1.5)--(6.5,1.5);
\draw[->,thick] (5.9,.6)--(6.5,.6);
\draw[->,thick] (5.9,-.3)--(6.5,-.3);
\node at (7.3, 1.5) {\scriptsize $\hat{m}_3, \hat{m}_2, \hat{m}_1$};
\node at (7.3, .6) {\scriptsize $\hat{\hat{m}}_3, \hat{\hat{m}}_2$};
\node at (7.3, -.3) {\scriptsize $\hat{\hat{\hat{m}}}_3$};
\end{tikzpicture}
\caption{Three-receiver less noisy broadcast channel with side information.}
\label{Fig2}
\end{figure}
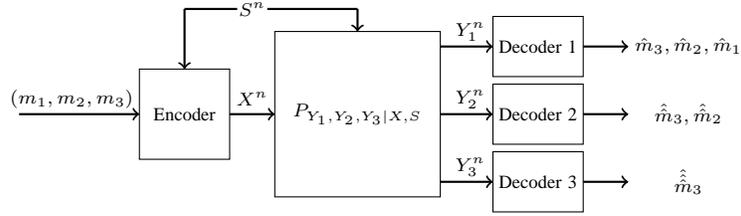

\textbf{Definition 4}:
In Fig. \ref{Fig2}, $3$-receiver less noisy BC with SI is depicted. Here according to \cite{nairwang2011}, we have $Y_3 \preccurlyeq Y_2 \preccurlyeq Y_1$ i.e. $Y_1$ is less noisy than $Y_2$ and $Y_2$ is less noisy than $Y_3$. 
As it is shown in Fig. \ref{Fig2}, we have a transmitter with $3$-degraded message sets $\left(\mathcal{M}_1, \mathcal{M}_2, \mathcal{M}_3 \right)$ which must be sent to three receivers. Here $m_3$ is the common message and is sent to all the receivers. The message $m_2 \in \mathcal{M}_2$ is sent to receivers $Y_1$ and $Y_2$ and finally the message $m_1 \in \mathcal{M}_1$ will be sent only for the receiver $Y_1$.

The definitions of code and rate tuple are as below:
{\small
\begin{equation}
\left(n, 2^{nR_1}, 2^{nR_2}, 2^{nR_3}, \epsilon \right) \nonumber \\
\end{equation}}
{\small
\begin{equation}
\left(R_1, R_2, R_3 \right)=\frac{1}{n}\left(\log M_1, \log M_2, \log M_3 \right). \nonumber 
\end{equation}}

The definitions of achievable rate tuples and the capacity region are similar to MBC.

\textbf{Definition 5}:
The model of fading Gaussian $3$-receiver less noisy broadcast channel is as follows:
{\small
\begin{eqnarray}
\label{fading-model}
Y_1= H_1X+Z_1 \nonumber \\
Y_2= H_2X+Z_2 \nonumber \\
Y_3= H_3X+Z_3.  
\end{eqnarray}}

For this fading channel, fading processes are ergodic stationary processes and we have the fading coefficients $H_1, H_2$ and $H_3$ related to receivers $Y_1, Y_2$ and $Y_3$, respectively. So we have the channel state as $\textbf{S}=\left(H_1,H_2, H_3 \right)$. We assume this channel state is available at the receivers perfectly and at the transmitter partially. For our fading Gaussian channel we consider three noises $Z_1, Z_2$ and $Z_3$ which are independent complex Gaussian random variables with zero means and unit variances related to three i.i.d. processes belonging to the path to each receiver and are independent of the channel state and auxiliary random variables. 

\textbf{Definition 6}:  
We show the partial channel state information at the transmitter (CSIT) by a deterministic function $f\left( . \right )$ as follows:
{\small
\begin{equation}
f\left(. \right): \mathcal{C}^3 \rightarrow \mathcal{A}, \nonumber 
\end{equation}}
where $\mathcal{A}$ is an arbitrary set and $\mathcal{C}$ denotes the complex number set. So we represent partial CSIT as $K=f\left(\textbf{S} \right)$.

\textbf{Definition 7}:
For the fading Gaussian $3$-receiver less noisy BC with messages $m_1$, $m_2$ and $m_3$, sent according to what mentioned in definition 4, in which the state information is available at the transmitter partially and at the receivers perfectly, a $\left(n,2^{nR_1}, 2^{nR_2}, 2^{nR_3},\epsilon\right)$ code, consists of an encoder map as below:
{\small
\begin{equation}
e_i: \{1,2,\cdots ,M_1\}\times \{1,2,\cdots ,M_2\} \times \{1,2,\cdots ,M_3\} \times \mathcal{A} \rightarrow \mathcal{C}, \nonumber
\end{equation}}
in which the signals are produced as $X_i=e_i\left( m_1,m_2,m_3, K_i\right)$, where $K_i=f\left(\textbf{S}_i \right)$ and $\textbf{S}_i$ is current channel state, $i=1,\cdots,n$, and we have average power constraint at the transmitter as follows:
{\small
\begin{equation}
\frac{1}{n}E\{ \sum_{i=1}^n\mid X_i\mid^2\}\leq P, \nonumber
\end{equation}}
and consists of decoding maps as below:
{\small
\begin{eqnarray}
d_{y1}: \mathcal{Y}_1^n, \mathbf{S}^n \rightarrow \{1,2,\cdots ,M_1 \} \times \{1,2,\cdots ,M_2\} \times \{1,2,\cdots ,M_3\} \nonumber\\
d_{y2}: \mathcal{Y}_2^n,  \mathbf{S}^n \rightarrow \{1,2,\cdots ,M_2\}\times \{1,2,\cdots ,M_3\} \nonumber\\
d_{y3}: \mathcal{Y}_3^n,  \mathbf{S}^n \rightarrow \{1,2,\cdots ,M_3\}, \nonumber
\end{eqnarray}}
such that $P_e^{(n)}\leq \epsilon$ , i.e.:

{\small
\begin{IEEEeqnarray}{ll}
\frac{1}{M_1M_2M_3}\sum_{m_1=1}^{M_1}\sum_{m_2=1}^{M_2} \sum_{m_3=1}^{M_3} \sum_{s^n\in \mathcal{S}^n} p\left(s^n\right)p\{d_{y1}\left(y_1^n,\mathbf{s}^n\right)\neq \left(m_1, m_2,m_3\right) \nonumber \\~or~d_{y2}\left(y_2^n, \mathbf{s}^n\right)\neq \left(m_2,m_3\right) ~or~ d_{y3}\left(y_3^n, \mathbf{s}^n\right)\neq m_3 \mid \mathbf{s}^n,x^n \left( m_1,m_2,m_3,k^n\right)\}\leq \epsilon. \nonumber
\end{IEEEeqnarray}}
The rate triplet of the code is denoted by $\left( R_1,R_2,R_3\right)$. If for any $\xi>0$ there exists an integer $n_0$ such that for all $n\geq n_0$ there is a $\left( n,2^{n\left( R_1-\xi\right)},2^{n\left( R_2-\xi\right)},2^{n\left( R_3-\xi\right)},\epsilon \right)$ code, then a rate triplet $\left( R_1,R_2,R_3\right)$ is $\epsilon$-achievable. The capacity region is the union of the closure of all $\epsilon$-achievable rate triplets.

\section{Multilevel Broadcast Channel with Non-causal Side Information}
In this section, first we derive achievable rate region for discrete memoryless MBC with non-causal SI available only at the transmitter then, we obtain capacity outer bound for this channel when non-causal SI is available at both transmitter and receivers .
\subsection{Achievable rate region when non-causal SI is available only at the transmitter}
Here, we obtain achievable rate region for discrete memoryless MBC with non-causal SI available at the transmitter.

Let $\mathcal{P}$ be the collection of all random variables $\left(U, V, W, S, X, Y_1, Y_2, Y_3 \right)$ with finite alphabets such that:
{\small
\begin{equation}
\label{MBC-distribution}
p\left(u, v, w, s, x, y_1, y_2, y_3 \right)= p\left(s \right) p\left(u\mid s \right) p\left( v\mid u,s \right)p\left(w\mid u,v,s\right)p\left(x\mid u,v,w,s \right) p\left(y_1, y_3\mid x,s\right)p\left(y_2\mid y_1\right).
\end{equation}}

\begin{theorem}
A pair $\left(R_0,R_1\right)$ is achievable for discrete memoryless MBC with SI non-causally available at the transmitter if we have:
{\small
\begin{IEEEeqnarray}{lll} 
\label{MBC-achievableRegion}
R_0 \leq \min \{ I\left(U;Y_2 \right)-I\left(U;S\right), I\left(UV; Y_3\right)-I\left(UV;S\right) \} \nonumber \\
R_1 \leq \min \{ I\left(VW;Y_1\mid U\right)-I\left(VW;S\mid U \right), I\left(W;Y_1\mid UV\right)+I\left(V;Y_3\mid U\right)-I\left(VW;S\mid U \right) \} \nonumber \\
R_0+R_1 \leq \min \{ I\left(UVW;Y_1\right)-I\left(UVW;S\right), I\left(W;Y_1\mid UV\right) +I\left(UV;Y_3\right)-I\left(UVW;S\right) \},
\end{IEEEeqnarray}}
for some $\left( U, V, W, S, X, Y_1, Y_2, Y_3 \right) \in \mathcal{P}$.
\end{theorem} 
  
\textbf{Corollary 1.1:}
Set $S=\phi$ and put $W=X$ in $\left(\ref{MBC-achievableRegion}\right)$, also consider the distribution in \cite{nairelg2009} for MBC, i.e.,
{\small
\begin{equation}
p\left(u, v, x, y_1, y_2, y_3 \right)= p\left(u\right) p\left( v\mid u\right)p\left(x\mid v\right)p\left(y_1, y_3\mid x\right)p\left(y_2\mid y_1\right), \nonumber
\end{equation}}
then this region reduces to achievable part of capacity theorem of MBC without SI \cite{nairelg2009}. First, we obtain following inequalities:
{\small
\begin{eqnarray}
R_0\leq \min \{ I\left(U;Y_2\right), I\left(V;Y_3\right) \} \nonumber \\
\label{S=tohi-MBCacheiv}
R_1\leq \min \{ I\left(X;Y_1\mid U\right), I\left(X;Y_1\mid V\right)+I\left(V;Y_3\mid U\right) \} \\
R_0+R_1\leq \min \{ I\left(X;Y_1\right), I\left(X;Y_1\mid V\right)+I\left(V;Y_3\right)\}. \nonumber  
\end{eqnarray}}
 From the argument mentioned in \cite{nairelg2009}, for second inequality we just have the first term i.e.:
{\small
 \begin{equation}
 R_1\leq I\left(X;Y_1\mid U\right), \nonumber
 \end{equation}}
consequently, for the third inequality the term $I\left(X;Y_1\right)$ is redundant, because:
\begin{itemize}
\item if: $I\left(U;Y_2\right) \leq I\left(V;Y_3\right)$: 
by adding both sides of inequalities related to $R_0$ and $R_1$ in $\left(\ref{S=tohi-MBCacheiv}\right)$ and according to the fact that $Y_2$ is a degraded version of $Y_1$ it is obvious that $R_0+R_1\leq I\left(X;Y_1\right)$.

\item if: $I\left(U;Y_2\right) \geq I\left(V;Y_3\right)$: 
because $Y_2$ is a degraded version of $Y_1$ we have:
{\small
\begin{equation}
I\left(U;Y_2 \right) \leq I\left(U;Y_1\right), \nonumber
\end{equation}}
consequently we have:
{\small
\begin{equation}
I\left(V;Y_3\right)\leq I\left(U;Y_1\right), \nonumber
\end{equation}}
therefore, again, from adding both sides of inequalities related to $R_0$ and $R_1$ in $\left(\ref{S=tohi-MBCacheiv}\right)$ it can be easily seen that $R_0+R_1\leq I\left(X;Y_1\right)$.
\end{itemize}
 So the rate region $\left( \ref{MBC-achievableRegion}\right)$ can be written as below which is identical to the capacity of MBC without SI \cite{nairelg2009}:
{\small
 \begin{eqnarray}
 R_0\leq \min \{ I\left(U;Y_2\right), I\left(V;Y_3\right) \} \nonumber \\
 R_1\leq I\left(X;Y_1\mid U\right)\nonumber \\
 R_0+R_1\leq I\left(X;Y_1\mid V\right)+I\left(V;Y_3\right), \nonumber 
 \end{eqnarray}}
 for some $p\left(u, v, x, y_1, y_2, y_3 \right)= p\left(u\right) p\left( v\mid u\right)p\left(x\mid v\right)p\left(y_1, y_3\mid x\right)p\left(y_2\mid y_1\right)$. 
 
\textbf{Corollary 1.2:}
By setting $V=U$ and $Y_3=Y_2$ in $\left( \ref{MBC-achievableRegion}\right)$, this region reduces to achievable rate region of 2-receiver degraded BC with non-causal SI at the encoder \cite{steinberg2005}. Let us show this result. The first inequality of $\left( \ref{MBC-achievableRegion}\right)$ is:
{\small
\begin{equation}
R_0\leq I\left(U;Y_2\right)-I\left(U;S\right), \nonumber 
\end{equation}}
also the second inequality of $\left( \ref{MBC-achievableRegion}\right)$ is:
{\small
\begin{equation}
R_1\leq I\left(W;Y_1\mid U\right)-I\left(W;S\mid U\right), \nonumber  
\end{equation}}
and for the last inequality of $\left( \ref{MBC-achievableRegion}\right)$ according to the fact that $Y_2$ is a degraded version of $Y_1$ we have following redundant inequality which can be derived from adding above two iequalities:
{\small
\begin{equation}
R_0+R_1\leq I\left(W;Y_1\mid U\right)+I\left( U;Y_2\right)-I\left(UW;S\right). \nonumber  
\end{equation}}

\textbf{Corollary 1.3:} 
Achievable rate region for discrete memoryless MBC when we have non-causal SI available at both transmitter and receivers, is as follows:
{\small
\begin{eqnarray}
\label{MBC-with-SI-at-BOTHsides}
R_0\leq \min \{ I\left(U;Y_2\mid S\right), I\left(UV;Y_3\mid S\right) \} \nonumber \\
R_1\leq \min \{I\left(VW;Y_1\mid US\right), I\left(W;Y_1\mid UVS\right)+I\left(V;Y_3\mid US\right) \} \\
R_0+R_1\leq \min \{ I\left(UVW;Y_1\mid S\right), I\left(W;Y_1\mid UVS\right)+I\left(UV;Y_3\mid S\right) \} \nonumber
\end{eqnarray}}
for some $p\left(u, v, w, s, x, y_1, y_2, y_3 \right)= p\left(s\right)p\left(u\mid s\right) p\left( v\mid u, s\right)p\left(w\mid u, v, s\right)p\left(x\mid u, v, w, s\right)p\left(y_1, y_3\mid x, s\right)p\left(y_2\mid y_1\right)$. 

By replacing $Y_k$ with $\left(S,Y_k\right)$, $k=1,2,3$ in $\left(\ref{MBC-achievableRegion}\right)$,  $\left(\ref{MBC-with-SI-at-BOTHsides}\right)$ can be easily shown.

\begin{IEEEproof}
To obtain the region $\left( \ref{MBC-achievableRegion}\right)$ we use a combination of superposition coding, indirect decoding and Gel'fand-Pinsker coding scheme. Fix n and a joint distribution on $\mathcal{P}$. Let side information be distributed i.i.d. according to:
{\small
\begin{equation}
p\left( s^n \right)=\prod_{i=1}^n p_{_S}\left( s_i\right) , \nonumber \\
\end{equation}}
where $p_{_S}\left( .\right) $ denotes probability mass function on $\mathcal{S}$. At first, we split each message $m_1 \in \mathcal{M}_1$ into two independent sub-messages $m_{11}\in \mathcal{M}_{11}$ and $m_{12} \in \mathcal{M}_{12}$ so that $R_1=R_{11}+R_{12}$.

\textbf{Codebook Generation:}
For each message $m_0$ we produce a subcodebook, (or bin), consists of $2^{nR'_0}$ sequences $u^n\left( m_0, m'_0\right)$, $m_0\in \{1, 2, \cdots, 2^{nR_0} \}$, $m'_0 \in \{1, 2, \cdots, 2^{nR'_0}\}$, which are generated randomly and independently and each one i.i.d. according to $\prod_{i=1}^np_U\left(u_i\right)$. Now for each $u^n\left( m_0, m'_0 \right)$, generate randomly and independently $2^{n\left( R_{11}+R'_{11}\right)}$ sequences $v^n\left( m_0, m'_0, m_{11}, m'_{11}\right) $, $m_{11}\in \{1, 2, \cdots, 2^{nR_{11}}\}$, $m'_{11}\in \{1, 2, \cdots, 2^{nR'_{11}}\}$, each one i.i.d. according to $\prod_{i=1}^np_{V\mid U}\left(v_i\mid u_i \left(m_0, m'_0\right) \right)$.
And for each $u^n\left( m_0, m'_0 \right)$ and $v^n\left( m_0, m'_0, m_{11}, m'_{11} \right)$ generate randomly and independently $2^{n\left( R_{12}+R'_{12}\right)}$ sequences $w^n\left( m_0, m'_0, m_{11}, m'_{11}, m_{12}, m'_{12}\right) $, $m_{12}\in \{1, 2, \cdots, 2^{nR_{12}}\}$,  $m'_{12}\in \{1, 2, \cdots, 2^{nR'_{12}}\}$, each one i.i.d. according to $\prod_{i=1}^np_{W\mid UV}\left(w_i\mid u_i\left(m_0, m'_0\right), v_i\left(m_0, m'_0,m_{11}, m'_{11}\right)  \right)$. Subcodebooks and their codewords are provided for the transmitter and all the receivers. 

\textbf{Encoding:}
Note that our messages are subcodebook indices. We have the message pair $\left( m_0, m_1\right)$ and the side information $s^n$. We find $m_{11}$, and $m_{12}$. In the subcodebook $m_0$ of $u^n$ sequences look for a $m'_0$ such that the sequence $u^n$ is jointly typical with the given $s^n$ i.e. $\left(u^n\left(m_0, m'_0\right), s^n \right)\in \tau_\epsilon^{(n)}$. Then, in the subcodebook $m_{11}$ of $v^n$ sequences look for some $m'_{11}$ such that:
{\small
\begin{equation}
\left( u^n\left( m_0, m'_0 \right), v^n\left( m_0, m'_0, m_{11}, m'_{11}\right), s^n \right) \in \tau_\epsilon^{(n)}, \nonumber 
\end{equation}}
and finally, in the subcodebook $m_{12}$ of $w^n$ sequences look for some $m'_{12}$ such that:
{\small
\begin{equation}
\left( u^n\left( m_0, m'_0 \right), v^n\left(m_0, m'_0, m_{11}, m'_{11}\right) ,w^n \left(m_0, m'_0, m_{11}, m'_{11}, m_{12}, m'_{12}\right), s^n \right) \in \tau_\epsilon^{(n)}. \nonumber 
\end{equation}}
At the end, we trnsmit $x^n\left( u^n, v^n, w^n, s^n \right)$ which is generated according to $\prod_{i=1}^np_{X\mid UVWS}\left( x_i\mid u_i,v_i,w_i,s_i\right)$.

\textbf{Decoding:}
Consider, through the encoding procedure, the correct indices are found, i.e. $m'_0=1$, $m'_{11}=1$ and $m'_{12}=1$. The messages are uniformly distributed, so without loss of generality, assume that $\left( m_0, m_{11}, m_{12}\right) =\left(1,1,1\right)$ is sent.

By analyzing the second receiver $Y_2$, which receives $y_2^n$, there exists the following important error event:
{\small
\begin{equation}
E_2=\{ \left(u^n\left(m_0, m'_0 \right), y_2^n\right) \in \tau_\epsilon^{(n)} ~ for ~ some ~ m_0\neq 1 ~and ~m'_0 \neq 1 \}, \nonumber
\end{equation}}
 to not have decoding error, according to packing lemma, we must have the following inequality:
{\small
\begin{equation}
\label{DncodingErrorY_2}
R_0+R'_0\leq I\left(U;Y_2\right)- \delta_2 \left(\epsilon \right).  
\end{equation}}

Also, for the receiver $Y_1$, we have following important error events:
{\small
\begin{IEEEeqnarray}{lll}
E_{11}=\{ \left(u^n\left(1,1\right), v^n\left(1,1,1,1\right), w^n\left(1,1,1,1,m_{12}, m'_{12}\right), y_1^n \right) \in \tau_\epsilon^{(n)} ~for ~some ~m_{12}\neq 1~m'_{12}\neq 1 \}, \nonumber \\
E_{12}=\{ \left(u^n\left(1,1\right), v^n\left(1,1,m_{11},m'_{11}\right), w^n\left(1,1,m_{11},m'_{11},m_{12}, m'_{12}\right), y_1^n \right) \in \tau_\epsilon^{(n)} ~for~some~m_{11}\neq 1,m'_{11}\neq 1,m_{12}\neq 1~m'_{12}\neq 1 \}, \nonumber \\
E_{13}=\{ \left(u^n\left(m_0,m'_0\right), v^n\left(m_0,m'_0,m_{11},m'_{11}\right), w^n\left(m_0,m'_0,m_{11},m'_{11},m_{12}, m'_{12}\right), y_1^n \right) \in \tau_\epsilon^{(n)}\nonumber \\  
~~~~~~~~~~~~~~~~~~~~~~~~~~~~~~~~~~~~~~~~~~~~~~~~~~~~~~~~~~~~~~~~~~~~~~~~~~~~~~~~for~some~m_0\neq 1,m'_0\neq 1,m_{11}\neq 1,m'_{11}\neq 1,m_{12}\neq 1~m'_{12}\neq 1 \}, \nonumber 
\end{IEEEeqnarray}}
 here again to not have decoding error, according to above events and packing lemma, we must have:
 {\small
 \begin{eqnarray}
 R_{12}+R'_{12}&\leq I\left(W;Y_1\mid UV\right)-\delta_{11} \left(\epsilon\right) \nonumber \\
 \label{DecodingErrorY_1}
 R_{11}+R'_{11}+R_{12}+R'_{12}&\leq I\left(VW;Y_1\mid U\right)-\delta_{12} \left(\epsilon \right) \\
 R_0+R'_0+R_{11}+R'_{11}+R_{12}+R'_{12}&\leq I\left(UVW;Y_1\right)-\delta_{13}
 \left(\epsilon \right) .\nonumber  
 \end{eqnarray}}

Finally for the receiver $Y_3$ we have following important error events:
{\small
\begin{IEEEeqnarray}{ll}
E_{31}=\{ \left( u^n\left(1,1\right), v^n\left(1,1,m_{11},m'_{11}\right)\right)\in \tau_\epsilon^{(n)} ~for ~some ~m_{11}\neq 1~ and ~m'_{11}\neq 1 \}, \nonumber \\
E_{32}=\{ \left(u^n\left(m_0,m'_0\right), v^n\left(m_0,m'_0,m_{11},m'_{11}\right)\right) \in \tau_\epsilon^{(n)} ~for ~some ~m_0\neq 1, m'_0\neq 1, m_{11}\neq 1~ and ~ m'_{11}\neq 1 \}, \nonumber
\end{IEEEeqnarray}}
so we must have:
{\small
\begin{eqnarray}
R_{11}+R'_{11}\leq I\left(V;Y_3\mid U\right)-\delta_{31} \left(\epsilon \right) \nonumber \\
\label{DecodingErrorY_3}
R_0+R_0+R_{11}+R'_{11}\leq I\left(UV;Y_3\right)- \delta_{32} \left(\epsilon \right). 
\end{eqnarray}}

Also, to prevent encoding error, according to covering lemma we must have:
{\small
\begin{eqnarray}
I\left(U,S\right)+\delta'_2 \left(\epsilon \right)\leq R'_0 \nonumber \\
\label{Encoding Errors}
I\left(V;S\mid U\right)+\delta'_3 \left(\epsilon \right) \leq R'_{11} \\
I\left(W;S\mid UV\right)+\delta'_1 \left(\epsilon \right) \leq R'_{12}. \nonumber 
\end{eqnarray}}
In above inequalities, functions $\delta_2\left(\epsilon\right), \delta_{11}\left(\epsilon\right), \delta_{12}\left(\epsilon\right), \delta_{13}\left(\epsilon\right), \delta_{31}\left(\epsilon\right), \delta_{32}\left(\epsilon\right), \delta'_{1}\left(\epsilon\right), \delta'_{2}\left(\epsilon\right)$ and $\delta'_{3}\left(\epsilon\right)$ tend to zero as $\epsilon \rightarrow 0$. Now by combining $\left(\ref{DncodingErrorY_2}\right)-\left(\ref{Encoding Errors}\right)$ and using Fourier-Motzkin procedure we obtain $\left(\ref{MBC-achievableRegion}\right)$.
\end{IEEEproof}

\subsection{Capacity outer bound}
In this subsection we obtain a capacity outer bound for discrete memoryless MBC in presence of non-cauasal SI.
\begin{theorem}
A capacity outer bound for discrete memoryless MBC with non-causal SI available at both transmitter and receivers, is as follows:
{\small
\begin{eqnarray}
\label{MBC-converse}
R_0\leq \min \{ I\left(U;Y_2\mid S\right), I\left(UV;Y_3\mid S\right)\} \nonumber\\
R_1\leq \min \{ I\left(X;Y_1\mid US\right), I\left(X;Y_1\mid UVS\right)+I\left(V;Y_3\mid US\right)+I\left(U;Y_3\mid S\right) \} \\
R_0+R_1\leq \min \{ I\left(X;Y_1\mid S \right), I\left(X;Y_1\mid UVS\right)+I\left(UV;Y_3\mid S\right)+ I\left(U;Y_1\mid S\right) \}, \nonumber
\end{eqnarray}}
for some $p\left(u, v, w, s, x, y_1, y_2, y_3 \right)= p\left(s \right) p\left(u, v\mid s \right)p\left(w\mid u,v,s\right)p\left(x\mid u,v,w,s \right) p\left(y_1, y_3\mid x,s\right)p\left(y_2\mid y_1\right).$
\end{theorem}

\textbf{Remark 1:}
\label{remark 1}
Note that according to memorylessness of the channel we have the following Markov chain:
{\small
\begin{equation}
\label{memorylessMarkovity}
m_0, m_1, S^{i-1}, S_{i+1}^n, Y_1^{i-1}, Y_2^{i-1}, Y_{3 i+1}^n \rightarrow X_i,S_i \rightarrow Y_{1i}, Y_{2i}, Y_{3i}.  
\end{equation}}
Also we know in MBC, receiver $Y_2$ is a degraded version of receiver $Y_1$ so we have:
{\small
\begin{equation}
\label{DegradedMarkovity}
S,X \rightarrow Y_1 \rightarrow Y_2. 
\end{equation}}
If in $\left( \ref{MBC-with-SI-at-BOTHsides} \right)$ we consider $X=W$ then according to $\left( \ref{memorylessMarkovity} \right)$ we have following achievable rate region for discrete memoryless MBC with non-causal SI available at both transmitter and receivers:
{\small
\begin{eqnarray}
\label{MBC-W=X}
R_0\leq \min \{ I\left(U;Y_2\mid S\right), I\left(UV;Y_3\mid S \right) \} \nonumber \\
R_1\leq \min \{I\left(X;Y_1\mid US\right), I\left(X;Y_1\mid UVS\right)+I\left(V;Y_3\mid US\right) \} \\
R_0+R_1\leq \min \{ I\left(X;Y_1\mid S\right), I\left(X;Y_1\mid UVS\right)+I\left(UV;Y_3\mid S\right) \}, \nonumber
\end{eqnarray}}
for some $p\left(u, v, s, x, y_1, y_2, y_3 \right)= p\left(s\right)p\left(u\mid s\right) p\left( v\mid u, s\right)p\left(x\mid u, v, s\right)p\left(y_1, y_3\mid x, s\right)p\left(y_2\mid y_1\right)$.  

In this case the gap between  outer bound $\left(\ref{MBC-converse} \right)$, and inner bound $\left( \ref{MBC-W=X} \right)$, can be easily seen.

\textbf{Remark 2:}
In MBC, if $Y_3$ be less noisy than $Y_1$, in other word if we have two following conditions:
{\small
\begin{eqnarray}
I\left(V;Y_1\mid U S\right)\leq I\left(V;Y_3\mid U S\right) \nonumber \\
I\left(UV;Y_1\mid S\right)\leq I\left(UV;Y_3\mid S\right), \nonumber 
\end{eqnarray}}
then, achievable rate region $\left( \ref{MBC-W=X}\right)$ and outer bound $\left( \ref{MBC-converse}\right)$ coincide
and result in the following capacity region:
{\small
\begin{eqnarray}
R_0 \leq I\left(U;Y_2 \mid S\right) \nonumber \\
R_1 \leq I\left(X;Y_1\mid U S\right) \nonumber \\
R_0+R_1 \leq I\left(X;Y_1\mid S\right) \nonumber 
\end{eqnarray}}
for some $p\left( u, v, s, x, y_1, y_2, y_3 \right) = p\left( s\right) p\left( u\mid s\right) p\left( v\mid u,s\right) p\left( x\mid u,v,s\right) p\left(y_1,y_3\mid x,s\right) p\left( y_2\mid y_1\right)$. 

But this setting for the channel makes no sense because for the less noisy receiver than $Y_1$, i.e. $Y_3$, at the transmitter we decide to send less information than $Y_1$ i.e. only message $m_0$.  So this setting is disregarded.

\begin{IEEEproof}
Now we focus on proving $\left(\ref{MBC-converse}\right) $. Let $m_0$ and $m_1$ be random variables related to our messages. Now we start the proof. 
{\small
\begin{IEEEeqnarray}{lll}
\label{converR_0}
nR_0=H\left(m_0\right)=H\left(m_0\mid S^n\right)=H\left(m_0\mid Y_2^n, S^n\right)+I\left(m_0;Y_2^n\mid S^n\right) \mathop{\leq }^{(a)} n\epsilon_{1n}+\sum_{i=1}^nI\left(m_0;Y_{2i}\mid Y_2^{i-1}, S^n\right) \nonumber \\  
=n\epsilon_{1n}+\sum_{i=1}^n\left[ H\left(Y_{2i}\mid Y_2^{i-1}, S^n\right)-H\left(Y_{2i}\mid Y_2^{i-1}, m_0, S^n\right)\right] \leq n\epsilon_{1n}+\sum_{i=1}^n\left[ H\left(Y_{2i}\mid S_i\right)-H\left(Y_{2i}\mid Y_1^{i-1}, Y_2^{i-1}, S^n, m_0\right) \right] \nonumber \\
=n\epsilon_{1n}+\sum_{i=1}^n\left[ H\left(Y_{2i}\mid S_i\right)-H\left(Y_{2i}\mid m_0, Y_1^{i-1}, Y_2^{i-1}, S^{i-1}, S_{i+1}^n, S_i\right) \right]\mathop{=}^{(b)}n\epsilon_{1n}+\sum_{i=1}^nI\left(Y_{2i};U_i\mid S_i\right), 
\end{IEEEeqnarray}}
where $(a)$ is due to the Fano's inequality, used in capacity outer bound proofs throughout the paper, and $(b)$ results from setting $U_i\triangleq \left(m_0, Y_1^{i-1}, Y_2^{i-1}, S^{i-1}, S_{i+1}^n\right)$.

{\small
\begin{IEEEeqnarray}{lll}
\label{ConversR_0{2}}
nR_0=H\left(m_0\right)=H\left(m_0\mid S^n\right)=H\left(m_0\mid Y_3^n, S^n\right)+I\left(m_0;Y_3^n\mid S^n\right) \leq  n\epsilon_{2n}+\sum_{i=1}^nI\left(m_0;Y_{3i}\mid Y_{3 i+1}^{n}, S^n\right) \nonumber \\  
=n\epsilon_{2n}+\sum_{i=1}^n \left[ H\left(Y_{3i}\mid Y_{3 i+1}^{n}, S^n\right)-H\left(Y_{3i}\mid Y_{3 i+1}^{n}, m_0, S^n\right)\right] \leq n\epsilon_{2n}+\sum_{i=1}^n \left[ H\left(Y_{3i}\mid S_i\right)-H\left(Y_{3i}\mid Y_1^{i-1}, Y_2^{i-1}, Y_{3i+1}^n, S^n, m_0\right) \right] \nonumber \\
\mathop{=}^{(c)} n\epsilon_{2n}+\sum_{i=1}^nI\left(U_iV_i;Y_{3i}\mid S_i \right),  
\end{IEEEeqnarray}}
where $(c)$ results from setting $V_i\triangleq \left(m_0, Y_{3i+1}^n\right)$. 

{\small
\begin{IEEEeqnarray}{lll}
\label{ConversR_1}
nR_1=H\left(m_1\right)=H\left(m_1\mid S^n, m_0\right)=H\left(m_1\mid Y_1^n, Y_2^n,  S^n, m_0\right)+I\left(m_1;Y_1^n, Y_2^n\mid S^n, m_0\right) \nonumber \\ \leq  n\epsilon_{3n}+\sum_{i=1}^nI\left(m_1;Y_{1i}, Y_{2i}\mid Y_1^{ i-1}, Y_2^{i-1}, S^{i-1}, S_{i+1}^n, S_i, m_0\right)   
=n\epsilon_{3n}+\sum_{i=1}^n I\left(m_1;Y_{1i}, Y_{2i}\mid U_i, S_i\right) \nonumber \\ \mathop{\leq}^{(d)} n\epsilon_{3n}+\sum_{i=1}^n I\left(X_i;Y_{1i}, Y_{2i}\mid U_i, S_i\right)  \nonumber \\
=\sum_{i=1}^nI\left(X_i;Y_{1i}\mid U_i, S_i \right) + \sum_{i=1}^nI\left(X_i;Y_{2i}\mid Y_{1i}, U_i, S_i\right) \mathop{=}^{(e)} n\epsilon_{3n}+\sum_{i=1}^nI\left(X_i;Y_{1i}\mid U_i, S_i \right),  
\end{IEEEeqnarray}}
where $(d)$ is due to the data processing inequality and $(e)$ results from the fact that $Y_2$ is a degraded version of $Y_1$ i.e. $\left( \ref{DegradedMarkovity} \right)$.

{\small
\begin{IEEEeqnarray}{lll}
\label{ConversR_1{2}}
nR_1=H\left(m_1\right)=H\left(m_1\mid S^n, m_0\right)=H\left(m_1\mid Y_1^n, Y_2^n,  S^n, m_0\right)+I\left(m_1;Y_1^n, Y_2^n\mid S^n, m_0\right) \nonumber \\ \leq  n\epsilon_{3n}+\sum_{i=1}^nI\left(m_1;Y_{1i}, Y_{2i}\mid Y_1^{ i-1}, Y_2^{i-1}, S^n, m_0\right) \nonumber \\
=n\epsilon_{3n} +\sum_{i=1}^nI\left(m_1;Y_{1i}\mid Y_1^{ i-1}, Y_2^{i-1}, S^{i-1}, S_{i+1}^n, S_i, m_0\right) +\sum_{i=1}^nI\left(m_1;Y_{2i}\mid Y_{1i}, Y_1^{i-1}, Y_2^{i-1}, S^{i-1}, S_{i+1}^n, S_i, m_0 \right) \nonumber\\
\mathop{=}^{(f)} n\epsilon_{3n} +\sum_{i=1}^nI\left(m_1;Y_{1i}\mid Y_1^{ i-1}, Y_2^{i-1}, S^{i-1}, S_{i+1}^n, S_i, m_0\right) \nonumber\\  
\leq n\epsilon_{3n}+\sum_{i=1}^n I\left(m_1, X_i, S_i, Y_{3i+1}^n;Y_{1i}\mid m_0, Y_1^{i-1}, Y_2^{i-1}, S^{i-1}, S_{i+1}^n, S_i\right) \nonumber \\
=n\epsilon_{3n}+
\sum_{i=1}^n I\left(Y_{3i+1};Y_{1i}\mid m_0, Y_1^{i-1}, Y_2^{i-1}, S^{i-1}, S_{i+1}^n, S_i\right) +
\sum_{i=1}^n I\left(X_i;Y_{1i}\mid m_0, Y_1^{i-1}, Y_2^{i-1}, S^{i-1}, S_{i+1}^n, S_i, Y_{3i+1}^n\right) \nonumber \\
\mathop{=}^{(g)}n\epsilon_{3n}+ \sum_{i=1}^nI\left(Y_1^{i-1};Y_{3i}\mid m_0, Y_{3i+1}^n, Y_2^{i-1}, S^{i-1}, S_{i+1}^n ,S_i \right) + \sum_{i+1}^nI\left(X_i;Y_{1i}\mid U_i, V_i, S_i\right) \nonumber \\ \leq n\epsilon_{3n}+\sum_{i=1}^nI\left(m_0, Y_{3i+1}^n, Y_1^{i-1};Y_{3i}\mid m_0, Y_2^{i-1}, S^{i-1}, S_{i+1}^n, S_i \right)+\sum_{i=1}^nI\left(X_i; Y_{1i}\mid U_i, V_i, S_i\right) \nonumber \\
=n\epsilon_{3n}+\sum_{i=1}^n I\left( Y_1^{i-1}; Y_{3i}\mid m_0, Y_2^{i-1}, S^{i-1}, S_{i+1}^n, S_i \right) \nonumber \\
+ \sum_{i=1}^nI\left( m_0, Y_{3i+1}^n;Y_{3i}\mid m_0, Y_1^{i-1}, Y_2^{i-1}, S^{i-1}, S_{i+1}^n, S_i \right)+\sum_{i=1}^nI\left(X_i;Y_{1i}\mid U_i, V_i, S_i\right) \nonumber \\ 
\leq n\epsilon_{3n}+\sum_{i=1}^n I\left( m_0, Y_1^{i-1}, Y_2^{i-1}, S^{i-1}, S_{i+1}^n; Y_{3i}\mid S_i \right) + \sum_{i=1}^nI\left( V_i;Y_{3i}\mid U_i ,S_i \right)+\sum_{i=1}^nI\left(X_i;Y_{1i}\mid U_i, V_i, S_i\right) \nonumber \\
=n\epsilon_{3n}+\sum_{i=1}^n I\left(U_i;Y_{3i}\mid S_i\right) + \sum_{i=1}^nI\left( V_i;Y_{3i}\mid U_i ,S_i \right)+\sum_{i=1}^nI\left(X_i;Y_{1i}\mid U_i, V_i, S_i\right) , 
\end{IEEEeqnarray}}
where $(f)$  is with regard to the fact that $Y_2$ is a degraded version of $Y_1$ i.e. $\left( \ref{DegradedMarkovity}\right)$ and $(g)$ results from the Csisz\'{a}r sum identity.

{\small
\begin{IEEEeqnarray}{lll}
\label{ConversR_{0+1}}
n\left( R_0+R_1 \right)=H\left(m_0, m_1\right)=H\left(m_0, m_1\mid S^n\right)=H\left(m_0, m_1\mid S^n, Y_1^n\right)+I\left(m_0, m_1;Y_1^n\mid S^n\right) \nonumber \\ 
\leq  n\epsilon_{4n}+\sum_{i=1}^nI\left(m_0, m_1, X_i, S_i, S^{i-1}, S_{i+1}^n, Y_1^{i-1};Y_{1i}\mid S_i\right) \nonumber \\
\mathop{=}^{(h)}n\epsilon_{4n}+\sum_{i=1}^n \left(X_i;Y_{1i}\mid S_i \right),  
\end{IEEEeqnarray}}
where $(h)$ is due to the memorylessness property of the channel i.e. (\ref{memorylessMarkovity}).
 
{\small
\begin{IEEEeqnarray}{lll}
\label{ConversR_{0+1(2)}}
n\left( R_0+R_1 \right)=H\left(m_0, m_1\right)=H\left(m_0, m_1\mid S^n\right)=H\left(m_0, m_1\mid S^n, Y_1^n, Y_2^n\right)+I\left(m_0, m_1;Y_1^n, Y_2^n\mid S^n\right) \nonumber \\ 
\leq  n\epsilon_{5n}+\sum_{i=1}^nI\left(m_0, m_1;Y_{1i}, Y_{2i}\mid Y_1^{i-1}, Y_2^{i-1}, S^n\right) \nonumber \\
=n\epsilon_{5n}+\sum_{i=1}^nI\left(m_0, m_1; Y_{1i}\mid Y_1^{i-1}, Y_2^{i-1}, S^n \right)+\sum_{i=1}^nI\left(m_0, m_1; Y_{2i}\mid Y_{1i}, Y_1^{i-1}, Y_2^{i-1}, S^n \right) \nonumber \\
\mathop{=}^{(i)} n\epsilon_{5n}+\sum_{i=1}^n I\left(m_0, m_1;Y_{1i}\mid Y_1^{i-1}, Y_2^{i-1}, S^n \right) \nonumber \\
\leq n\epsilon_{5n}+\sum_{i=1}^nI\left(m_0, m_1, X_i, S_i, Y_{3i+1}^n;Y_{1i}\mid Y_1^{i-1}, Y_2^{i-1}, S^n \right) \nonumber \\
= n\epsilon_{5n}+\sum_{i=1}^nI\left(m_0, X_i, Y_{3i+1}^n;Y_{1i}\mid Y_1^{i-1}, Y_2^{i-1}, S^n \right) \nonumber \\
=n\epsilon_{5n}+ \sum_{i=1}^n I\left( m_0, Y_{3i+1}^n;Y_{1i}\mid Y_1^{i-1}, Y_2^{i-1}, S^n\right)+\sum_{i=1}^nI\left( X_i;Y_{1i}\mid Y_1^{i-1}, Y_2^{i-1}, Y_{3i+1}^n, S^n, m_0\right) \nonumber \\
=n\epsilon_{5n}+\sum_{i=1}^n I\left(m_0;Y_{1i}\mid Y_1^{i-1}, Y_2^{i-1}, S^n\right)+\sum_{i=1}^n
I\left(Y_{3i+1}^n;Y_{1i}\mid Y_1^{i-1}, Y_2^{i-1}, S^n, m_0 \right)+\sum_{i=1}^n I\left(X_i;Y_{1i}\mid U_i, V_i, S_i \right) \nonumber \\
\mathop{=}^{(j)} n\epsilon_{5n}+\sum_{i=1}^n I\left(m_0;Y_{1i}\mid Y_1^{i-1}, Y_2^{i-1}, S^n\right)+\sum_{i=1}^n
I\left(Y_1^{i-1};Y_{3i}\mid Y_{3i+1}^{n}, Y_2^{i-1}, S^n, m_0 \right)+\sum_{i=1}^n I\left(X_i;Y_{1i}\mid U_i, V_i, S_i \right) \nonumber \\
\leq n\epsilon_{5n}+\sum_{i=1}^n I\left(m_0, Y_1^{i-1}, Y_2^{i-1}, S^{i-1}, S_{i+1}^n; Y_{1i}\mid S_i\right) \nonumber \\
+\sum_{i=1}^nI\left( Y_1^{i-1}, Y_2^{i-1}, Y_{3i+1}^n, S^{i-1}, S_{i+1}^n, m_0;Y_{3i}\mid S_i \right) +\sum_{i=1}^n I\left(X_i;Y_{1i}\mid U_i, V_i, S_i \right) \nonumber \\
=n\epsilon_{5n}+\sum_{i=1}^n I\left( U_i; Y_{1i}\mid S_i \right)+ \sum_{i=1}^nI\left(U_iV_i;Y_{3i}\mid S_i\right) +\sum_{i=1}^n I\left(X_i;Y_{1i}\mid U_iV_iS_i\right),   
\end{IEEEeqnarray}}
where $(i)$ is due to the fact that $Y_2$ is a degraded version of $Y_1$ i.e. $\left( \ref{DegradedMarkovity}\right)$ and $(j)$ results from Csisz\'{a}r sum identity. 

Now according to $\left( \ref{converR_0}\right) -\left( \ref{ConversR_{0+1(2)}} \right)$ and by using the standard time-sharing scheme, it could be concluded that any achievable rate for the MBC with non-causal SI available at both the transmitter and receivers, must satisfy $\left( \ref{MBC-converse}\right)$.
\end{IEEEproof}

\section{3-Receiver Less Noisy Broadcast Channel with Non-causal Side Information}
In this section first we derive an achievable rate region for discrete memoryless $3$-receiver less noisy BC with non-causal SI available only at the transmitter then, we obtain capacity region for this channel when non-causal SI is available at both transmitter and receivers.
\subsection{Achievable rate region when non-causal SI is available only at the transmitter}
In this subsection we obtain an achievable rate region for $3$-receiver less noisy BC with non-causal SI available only at the transmitter. 

Let $\mathcal{P}^*$ be the collection of all random variables $\left(U, V, W, S, X, Y_1, Y_2, Y_3 \right)$ with finite alphabets such that:
{\small
\begin{equation}
\label{3-Less-noisy-distribution}
p\left(u, v, w, s, x, y_1, y_2, y_3 \right)= p\left(s \right) p\left(u\mid s \right) p\left( v\mid u,s \right)p\left(w\mid u,v,s\right)p\left(x\mid u,v,w,s \right) p\left(y_1,y_2,y_3\mid x,s\right). 
\end{equation}}

\begin{theorem}
A triplet of nonnegative numbers $\left(R_1,R_2,R_3\right)$ is achievable for discrete memoryless 3-receiver less noisy BC with SI non-causally available at the transmitter if we have:
{\small
\begin{IEEEeqnarray}{lll}
\label{3-less-noisy-BC-achievableRegion}
R_3 \leq I\left(U;Y_3 \right)-I\left(U;S\right) \nonumber \\
R_2 \leq I\left(V;Y_2\mid U\right)-I\left(V;S\mid U \right) \nonumber \\
R_1 \leq I\left(W;Y_1\mid UV\right)-I\left(W;S\mid UV\right), 
\end{IEEEeqnarray}}
for some $\left( U, V, W, S, X, Y_1, Y_2, Y_3 \right) \in \mathcal{P}^*$.
\end{theorem}

\textbf{Corollary 3.1:}
Set $S=\phi$ and put $W=X$ in $\left(\ref{3-less-noisy-BC-achievableRegion}\right)$, also consider following distribution for $3$-receiver less noisy BC, as in \cite{nairwang2011},
{\small
\begin{equation}
p\left(u, v, x, y_1, y_2, y_3 \right)= p\left(u\right) p\left( v\mid u\right)p\left(x\mid v\right)p\left(y_1, y_3\mid x\right)p\left(y_2\mid y_1\right), \nonumber
\end{equation}}
then, this region reduces to achievable part of capacity theorem of $3$-receiver less noisy BC without SI \cite{nairwang2011}. 

In addition to above replacement, by setting $V=U$ and $Y_2=Y_3$ in $\left(\ref{3-less-noisy-BC-achievableRegion}\right)$, then this region reduces to the achievable part of capacity theorem of $2$-receiver less noisy BC without SI.

\textbf{Corollary 3.2:}
Achievable rate region for $3$-receiver less noisy BC with non-causal SI available at both transmitter and receivers can be obtained by setting $Y_k=\left( S,Y_k\right)$, $k=1, 2, 3$, in $\left( \ref{3-less-noisy-BC-achievableRegion}\right)$ as follows:
{\small
\begin{IEEEeqnarray}{lll}
\label{3-less-noisy-bc-at-both-achiev}
R_3\leq I\left(U;Y_3\mid S\right) \nonumber \\
R_2\leq I\left(V;Y_2\mid U S\right) \nonumber \\
R_1\leq I\left( W;Y_1\mid U V S\right),
\end{IEEEeqnarray}}
for some $\left( U, V, W, S, X, Y_1, Y_2, Y_3 \right) \in \mathcal{P}^*$.
  
\begin{IEEEproof}
To obtain the region $\left( \ref{3-less-noisy-BC-achievableRegion} \right)$ we use a combination of superposition coding and Gel'fand-Pinsker coding scheme. The proof is similar to the proof of Theorem 1 and thus only an outline is presented. Fix n and a joint distribution on $\mathcal{P}^*$. Let side information be distributed i.i.d. according to:
{\small
\begin{equation}
p\left( s^n \right)=\prod_{i=1}^n p_{_S}\left( s_i\right). \nonumber 
\end{equation}}

For each message $m_3$ we produce a subcodebook (or bin) consists of $2^{nR'_3}$ sequences $u^n\left( m_3, m'_3\right)$, $m_3\in \{1, 2, \cdots, 2^{nR_3} \}$, $m'_3 \in \{1, 2, \cdots, 2^{nR'_3}\}$, which are generated randomly and independently and each one i.i.d. according to $\prod_{i=1}^np_U\left(u_i\right)$. Now for each $u^n\left( m_3, m'_3 \right)$, generate randomly and independently $2^{n\left( R_2+R'_2\right)}$ sequences $v^n\left( m_3, m'_3, m_2, m'_2\right) $, $m_2\in \{1, 2, \cdots, 2^{nR_2}\}$, $m'_2\in \{1, 2, \cdots, 2^{nR'_2}\}$, each one i.i.d. according to $\prod_{i=1}^np_{V\mid U}\left(v_i\mid u_i \left(m_3, m'_3\right) \right)$.
Next for each $u^n\left( m_3, m'_3 \right)$ and $v^n\left( m_3, m'_3, m_2, m'_2 \right)$ generate randomly and independently $2^{n\left( R_1+R'_1\right)}$ sequences $w^n\left( m_3, m'_3, m_2, m'_2, m_1, m'_1\right) $, $m_1\in \{1, 2, \cdots, 2^{nR_1}\}$, $m'_1\in \{1, 2, \cdots, 2^{nR'_1}\} $, each one i.i.d. according to $\prod_{i=1}^np_{W\mid UV}\left(w_i\mid u_i\left(m_3, m'_3\right), v_i\left(m_3, m'_3,m_2, m'_2\right)  \right)$. Subcodebooks and their codewords are provided for the transmitter and all receivers. 

Remind that our messages are subcodebook indices and we have the message triplet $\left( m_1, m_2, m_3\right)$ and the side information $s^n$. Now in the bin $m_3$ of $u^n$ sequences look for a $m'_3$ such that the sequence $u^n\left( m_3,m'_3\right)$ be jointly typical with the given $s^n$. Then in the bin $m_2$ of $v^n$ sequences look for some $m'_2$ such that:
{\small
\begin{equation}
\left( u^n\left( m_3, m'_3 \right), v^n\left( m_3, m'_3, m_2, m'_2\right), s^n \right) \in \tau_\epsilon^{(n)}, \nonumber 
\end{equation}}
and in the bin $m_1$ of $w^n$ sequences look for some $m'_1$ such that:
{\small
\begin{equation}
\left( u^n\left( m_3, m'_3 \right), v^n\left(m_3, m'_3, m_2, m'_2\right) ,w^n \left(m_3, m'_3, m_2, m'_2, m'_1, m_1\right), s^n \right) \in \tau_\epsilon^{(n)}. \nonumber 
\end{equation}}
Finally, we trnsmit $x^n\left( u^n, v^n, w^n, s^n \right)$ which is generated according to $\prod_{i=1}^np_{X\mid UVWS}\left( x_i\mid u_i,v_i,w_i,s_i\right)$.

If we have following inequalities, encoding will be successful with small probability of error:
{\small
\begin{IEEEeqnarray}{lll}
\label{encoding-3-rec-less-noisy-bc}
I\left( U;S\right) \leq R'_3 \nonumber \\
I\left( V;S\mid U\right) \leq R'_2 \nonumber \\
I\left(W;S\mid UV\right)\leq R'_1, 
\end{IEEEeqnarray}}
also decoding is successful if we have:
{\small
\begin{IEEEeqnarray}{lll}
\label{decoding-3-rec-less-noisy-bc}
R_3+R'_3\leq I\left( U;Y_3\right) \nonumber \\
R_2+R'_2 \leq I\left( V;Y_2\mid U\right) \nonumber \\
R_1+R'_1 \leq I\left(W;Y_1\mid UV\right) , 
\end{IEEEeqnarray}}
now by using Fourier-Motzkin procedure and $\left( \ref{encoding-3-rec-less-noisy-bc} \right)$-$\left( \ref{decoding-3-rec-less-noisy-bc} \right)$ we conclude $\left( \ref{3-less-noisy-BC-achievableRegion} \right)$. 
\end{IEEEproof}

\subsection{Capacity theorem of $3$-receiver less noisy BC with non-causal SI available at both transmitter and receivers}
Here, we obtain capacity region for discrete memoryless $3$-receiver less noisy BC with non-causal SI available at both transmitter and receivers.
\begin{theorem}
The capacity region of the $3$-receiver less noisy BC with non-causal SI available at both transmitter and receivers is the set of all rate triplets $\left(R_3,R_2,R_1\right)$ such that:
{\small
\begin{IEEEeqnarray}{lll}
\label{capacity-3-receiver-less-noisy-bc}
R_1\leq I\left( X;Y_1\mid U V S\right) \nonumber \\
R_2\leq I\left(V;Y_2\mid U S\right) \nonumber \\
R_3\leq I\left( U;Y_3\mid S\right), 
\end{IEEEeqnarray}}
for some $p\left( u,v,x,s,y_1,y_2,y_3\right) =p\left( s\right) p\left( u\mid s\right) p\left( v\mid u,s\right) p\left( x\mid u,v,s\right) p\left( y_1,y_2,y_3\mid x,s\right) $.
\end{theorem}

\begin{IEEEproof}

\textbf{Achievability:}
Achievable rate region can be easily obtained by setting $X=W$ in $\left( \ref{3-less-noisy-bc-at-both-achiev}\right)$.

\textbf{Converse:}
For converse proof we use an extension of lemma 1 in \cite{nairwang2011}. 

\textbf{Lemma:} Let the channel from $X$ to $Y$ be less noisy than $X$ to $Z$ in presence of SI. Consider $\left(M\right)$ to be any random variable such that:
{\small
\begin{equation}
M\rightarrow X^n, S^n\rightarrow  Y^n, Z^n \nonumber
\end{equation}}
forms a Markov chain, then we have:
{\small
\begin{eqnarray}
I\left(Y^{i-1};Y_i\mid M, S^n\right) \geq I\left( Z^{i-1};Y_i\mid M, S^n\right) \nonumber \\
I\left(Y^{i-1};Z_i\mid M, S^n\right) \geq I\left( Z^{i-1};Z_i\mid M, S^n\right), \nonumber
\end{eqnarray}}
where $1\leq i\leq n$.

Proof of the lemma is identical to what mentioned in \cite{hajizadehhodtani} and is ignored.
  
%
Now we start to prove the converse part. Let $m_1$, $m_2$ and $m_3$ be random variables related to our messages. 
{\small
\begin{IEEEeqnarray}{lllllll}
\label{R_3-3-less-noisy-bc}
nR_3=H\left(m_3\right)=H\left(m_3\mid S^n\right)=H\left(m_3\mid S^n, Y_3^n\right)+I\left(m_3;Y_3^n\mid S^n\right) \nonumber \\
\leq n\epsilon_{3n}+\sum_{i=1}^nI\left(m_3;Y_{3i}\mid Y_3^{i-1},S^n\right) \nonumber \\
\leq n\epsilon_{3n}+\sum_{i=1}^nI\left(m_3,S^{i-1},S_{i+1}^n,Y_3^{i-1};Y_{3i}\mid S_i \right) \nonumber \\
\mathop{\leq}^{(a)} n\epsilon_{3n}+\sum_{i=1}^nI\left(m_3,S^{i-1},S_{i+1}^n,Y_2^{i-1};Y_{3i}\mid S_i\right) \nonumber \\
= n\epsilon_{3n}+\sum_{i=1}^nI\left(U_i;Y_{3i}\mid S_i\right), 
\end{IEEEeqnarray}}
where $(a)$ results from the lemma. Also we have $U_i\triangleq \left(m_3,S^{i-1},S_{i+1}^n,Y_2^{i-1}\right)$. 

{\small
\begin{IEEEeqnarray}{lllll}
\label{R_2-3-less-noisy-bc}
nR_2=H\left( m_2\right) = H\left(m_2\mid m_3,S^n\right)=H\left(m_2\mid m_3,S^n,Y_2^n\right)+I\left(m_2;Y_2^n\mid m_3,S^n\right) \nonumber \\
\leq n\epsilon_{2n}+\sum_{i=1}^nI\left( m_2;Y_{2i}\mid Y_2^{i-1},m_3,S^n\right) \nonumber \\
=n\epsilon_{2n}+\sum_{i=1}^nI\left( m_2,m_3,Y_2^{i-1};Y_{2i}\mid Y_2^{i-1}, m_3,S^n\right) \nonumber \\ 
=n\epsilon_{2n}+\sum_{i=1}^nI\left( V_i;Y_{2i}\mid U_i,S_i\right),
\end{IEEEeqnarray}}
where $V_i\triangleq \left(m_2,m_3,Y_2^{i-1}\right)$.

{\small
\begin{IEEEeqnarray}{llllllll} 
\label{R_1-3-less-noisy-bc}
nR_1=H\left( m_1\right) =H\left(m_1\mid m_2,m_3,S^n\right)=H\left(m_1\mid m_2,m_3,S^n,Y_1^n\right)+I\left( m_1;Y_1^n\mid m_2,m_3,S^n\right) \nonumber \\
\leq n\epsilon_{1n}+\sum_{i=1}^nI\left(m_1;Y_{1i}\mid Y_1^{i-1},m_2,m_3,S^n\right) \nonumber \\
\mathop{\leq}^{(b)}n\epsilon_{1n}+\sum_{i=1}^nI\left(X_i;Y_{1i}\mid Y_1^{i-1},m_2,m_3,S^n\right) \nonumber \\
=n\epsilon_{1n}+\sum_{i=1}^nI\left(X_i;Y_{1i}\mid m_2,m_3,S^n\right) -\sum_{i=1}^nI\left(Y_1^{i-1};Y_{1i}\mid m_2,m_3,S^n\right) \nonumber \\
\mathop{\leq}^{(c)}n\epsilon_{1n}+\sum_{i=1}^nI\left(X_i;Y_{1i}\mid m_2,m_3,S^n\right) -\sum_{i=1}^nI\left(Y_2^{i-1};Y_{1i}\mid m_2,m_3,S^n\right) \nonumber \\
=n\epsilon_{1n}+\sum_{i=1}^nI\left(X_i;Y_{1i}\mid ,Y_2^{i-1}m_2,m_3,S^n\right) \nonumber \\
=n\epsilon_{1n}+\sum_{i=1}^nI\left(X_i;Y_{1i}\mid U_iV_iS_i\right), 
\end{IEEEeqnarray}}
where $(b)$ and $(c)$ result from the data processing inequality and the lemma, respectively. Now according to $\left( \ref{R_3-3-less-noisy-bc}\right),\left( \ref{R_2-3-less-noisy-bc}\right) ,\left( \ref{R_1-3-less-noisy-bc}\right)$ and by using the standard time-sharing scheme, capacity outer bound for discrete memoryless $3$-receiver less noisy BC is obtained as $\left(\ref{capacity-3-receiver-less-noisy-bc}\right)$.
\end{IEEEproof}

\section{Fading Gaussian 3-Receiver Less Noisy Broadcast Channel with Partial CSIT}
In this section we derive capacity bounds for fading Gaussian $3$-receiver less noisy BC using our capacity theorem of discrete alphabet $3$-receiver less noisy BC with non-causal SI. Note that we assume our capacity theorem for discrete alphabet channel can be extended to discrete time and continuous alphabet fading Gaussian channel according to the mathematical point stated in \cite{gallagerbook}. We consider SI as channel state i.e. fading coefficients and the channel state is available partially at the transmitter and perfectly at the receivers. Considering partial CSIT is a practical consideration due to the fact that often perfect CSIT is not known. Here, we know since there is state information at the transmitter we need a power allocation function, $\varphi \left( .\right)$, defined as follows:
{\small
\begin{equation}
\varphi \left( . \right) : \mathcal{A} \rightarrow R_+, \nonumber
\end{equation}}
where $\mathcal{A}$, as mentioned in definition $6$, is an arbitrary set and $R_+$ denotes non-negative real numbers. Also, we define two deterministic functions $\alpha\left( .\right)$ and $\beta \left( . \right)$ in acheivable rate region proof for illustrating the power shares of each message and are defined as follows:
{\small
\begin{equation}
\alpha \left( . \right)~and~\beta \left( . \right) :\mathcal{A} \rightarrow \left[0,1\right] ,  \nonumber
\end{equation}}
also, we use two other deterministic functions $\vartheta \left( .\right)$ and $\gamma \left( . \right)$ in converse part proof which are defined as below:
{\small
\begin{equation}
\vartheta \left( . \right)~and~\gamma \left( . \right) :\mathcal{C}^3 \rightarrow \left[0,1\right]. \nonumber
\end{equation}}
Since here there is CSIT we have expexted average power constraint, i.e. we have $E\{\varphi\left(K\right)\}\leq P$. Note that $E\{ . \}$ denotes expectation operator with respect to the distribution on the channel state and $K$ represents partial CSIT. 

At the following subsections first we derive the capacity inner bound for the mentioned fading Gaussian $3$-receiver less noisy BC then, capacity outer bound for this channel is obtained and finally we show that for the special case in which we have perfect CSIT these two bounds coincide under certain conditions. 

\subsection{Inner bound for the fading Gaussian 3-receiver less noisy broadcast channel}
In this subsection we introduce an efficient signaling scheme to obtain achievable rate region for the fading Gaussian 3-receiver less noisy BC. 

\begin{theorem}
For a fading Gaussian $3$-receiver less noisy broadcast channel $\left( \ref{fading-model}\right)$, triplet $\left( R_1, R_2, R_3\right) \in R_+^3$ is achievable if we have:
{\small
\begin{IEEEeqnarray}{lll}
\label{achiev-Gaussian-fadin-3-less}
R_1\leq E\{\psi \left[\mid H_1\mid^2\varphi\left( K\right)\bar{ \alpha }\left( K\right) \right] \} \nonumber \\
R_2\leq E\{\psi \left[ \frac{\mid H_2\mid^2\varphi\left( K\right) \alpha \left( K\right) \bar{\beta}\left( K\right)}{\mid H_2\mid^2\varphi\left( K\right)\bar{ \alpha }\left( K\right) +1} \right] \} \nonumber \\
R_3\leq E\{ -\psi \left[ \frac{-\mid H_3\mid^2\varphi\left( K\right) \alpha \left( K\right) \beta\left( K\right)}{\mid H_3\mid^2\varphi\left( K\right) +1} \right] \}, 
\end{IEEEeqnarray}}
for all arbitrary $\alpha \left( . \right)$ and $\beta \left( . \right)$ which are deterministic functions from $\mathcal{A}$ to $\left[0,1\right]$ and for all power allocation functions $\varphi \left( . \right)$ from $\mathcal{A}$ to $R_+$ so that $E\{ \varphi \left( K \right) \}\leq P$.
\end{theorem}

\begin{IEEEproof}
To obtain the achievable rate region for this fading Gaussian channel we use the capacity region obtained in this paper for discrete alphabet 3-receiver less noisy BC with non-causal SI available at both transmitter and receivers $\left( \ref{capacity-3-receiver-less-noisy-bc}\right)$. First let us define an appropriate signaling. Auxiliary random variable $U$ is related only to message $m_3$, so let us define it as below:
{\small
\begin{equation}
U\triangleq \sqrt{\alpha \left( K\right)}U'. \nonumber 
\end{equation}}
Also we know auxiliary random variable $V$ is related to messages $m_2$ and $m_3$. We define it as follows:
{\small
\begin{equation}
V\triangleq \sqrt{\beta \left(K \right)}U'+\sqrt{\bar{\beta} \left(K\right)}V', \nonumber
\end{equation}}
and we know that random variable $X$ indicates messages $m_1$,$m_2$ and $m_3$. We define it as below:
{\small
\begin{equation}
\label{Signaling}
X\triangleq \sqrt{\varphi \left( K\right)}\left( \sqrt{\alpha \left( K\right)} V+\sqrt{\bar{\alpha}\left(K \right)}X'\right)
=\sqrt{\varphi \left(K \right)}\left( \sqrt{\alpha \left( K\right) \beta \left( K\right)} U'+ \sqrt{\alpha \left( K\right) \bar{\beta} \left( K\right)} V'+\sqrt{\bar{\alpha} \left( K \right)}X'\right).
\end{equation}}
Here, $U'$, $V'$ and $X'$ are independent Gaussian random variables with zero means and unit variances where each of them denotes the message $m_3, m_2$ and $m_1$, respectively and all of them are independent of the channel state. Here, as mentioned earlier $\alpha\left( .\right)$ and $\beta\left( .\right)$ are two deterministic functions from the set $\mathcal{A}$ to the interval $\left[0,1\right]$. As we see, expected average power constraint is satisfied.

Now let us return to inequalities in $\left( \ref{achiev-Gaussian-fadin-3-less} \right)$. For the first inequality from $\left( \ref{capacity-3-receiver-less-noisy-bc}\right)$ we have:
{\small
\begin{equation}
\label{R_1-Gaussian-fading}
R_1\leq I\left( X;Y_1\mid UV \mathbf{S}\right)=\int_\mathbf{s}I\left(X;Y_1\mid UV,\mathbf{S}=\mathbf{s}\right)f\left( \mathbf{s}\right)d\mathbf{s}, 
\end{equation}}
where, as described earlier $\mathbf{S}=\left( H_1, H_2, H_3\right)$ is channel state and $f\left( \mathbf{s} \right)$ denotes the channel state distribution. Let us evaluate $I\left(X;Y_1\mid UV,\mathbf{S}=\mathbf{s}\right)$ in the above expression:
{\small
\begin{IEEEeqnarray}{ll}
I\left(X;Y_1\mid UV,\mathbf{S}=\mathbf{s}\right)=H\left(Y_1\mid UV,\mathbf{S}=\mathbf{s}\right)-H\left(Y_1\mid UVX,\mathbf{S}=\mathbf{s}\right) \nonumber \\
=H\left(Y_1\mid UV,\mathbf{S}=\mathbf{s}\right)-\log \pi e.
\end{IEEEeqnarray}}
The first expression is equal to:
{\small
\begin{IEEEeqnarray}{ll} 
\label{R_1+Gauss-fad}
H\left( Y_1\mid UV,\mathbf{S}=\mathbf{s}\right) = \nonumber \\ H\left(h_1\left[\sqrt{\varphi\left( k\right)}\left(\sqrt{\alpha \left( k\right) \beta \left( k\right)}U'+\sqrt{\alpha \left( k\right) \bar{\beta} \left( k\right)}V'+\sqrt{\bar{\alpha} \left( k\right)}X' \right) \right]+Z_1 \mid \sqrt{\alpha \left( k\right)} U',\left( \sqrt{\beta \left( k\right)}U'+\sqrt{\bar{\beta} \left( k\right)}V' \right),\mathbf{S}=\mathbf{s}\right) \nonumber \\
=\log \pi e\left( \mid h_1\mid^2 \varphi \left( k\right) \bar{\alpha}\left( k\right) +1\right).
\end{IEEEeqnarray}}
From (\ref{R_1-Gaussian-fading}) - (\ref{R_1+Gauss-fad}) we can derive:
{\small
\begin{equation}
R_1\leq E\{\psi \left[ \mid H_1\mid^2 \varphi \left(K \right) \bar{\alpha}\left( K\right) \right] \}. \nonumber
\end{equation}}
The second and third inequalities in (\ref{achiev-Gaussian-fadin-3-less}) can be obtained in a similar way to the first one but let us show the third inequality more explicitly. From (\ref{capacity-3-receiver-less-noisy-bc}) we have:
{\small
\begin{equation}
R_3\leq I\left( U;Y_3\mid \mathbf{S}\right)=\int_\mathbf{s}I\left( U;Y_3\mid \mathbf{S}=\mathbf{s}\right) f\left( \mathbf{s}\right) d\mathbf{s}, \nonumber
\end{equation}}
and we have:
{\small
\begin{IEEEeqnarray}{ll}
\label{8-maghale khodam}
I\left( U;Y_3\mid \mathbf{S}=\mathbf{s}\right)=H\left( Y_3\mid \mathbf{S}=\mathbf{s}\right)-H\left( Y_3\mid U,\mathbf{S}=\mathbf{s}\right) \nonumber \\
=\log \pi e\left( \mid h_3\mid^2 \varphi \left( k\right) +1\right) -\log \pi e \left( \mid h_3 \mid^2 \varphi \left( k\right) \left[ \alpha \left( k\right) \bar{\beta}\left( k\right)+\bar{\alpha}\left( k\right) \right] +1\right).
\end{IEEEeqnarray}}
Since $\alpha \left( k\right) \bar{\beta}\left( k\right)+\bar{\alpha}\left( k\right)=1-\alpha \left( k\right) \beta \left( k\right)$ the third inequality is obtained.
\end{IEEEproof}

\textbf{Remark 3:}
Note that if we had considered a general 3-receiver BC with three degraded message sets, we would have noted some messages are not sent to some receivers, therefore according to the signaling $\left( \ref{Signaling}\right)$ and the role of each auxiliary random variable in it and also the fact that we have three independent noises with unit variances and zero means, we see when we have $\mid h_3\mid > \mid h_1\mid $ then, receiver $Y_1$ is a degraded version of receiver $Y_3$ and so the messages $m_1$ and $m_2$ could not be received by $Y_1$. Also when  $\mid h_3\mid > \mid h_2\mid $ then, receiver $Y_2$ is a degraded version of $Y_3$ and the receiver $Y_2$ can not receive message $m_2$. So from our signaling scheme it is clear that the function $\bar{\beta}\left( K\right)$ must have a form as below:
{\small
\begin{align}
\bar{\beta}\left( K\right) \triangleq \left \{
\begin{matrix}
\bar{\beta}^{*} \left( K\right)  & if: \mid H_1\mid > \mid H_3\mid or  \mid H_2\mid > \mid H_3\mid \\ 0 & otherwise,
\end{matrix} \right.
\nonumber
\end{align}}
in which $\bar{\beta}^{*} \left( .\right)$ is a deterministic function from $\mathcal{A}$ to the interval $\left[0,1\right]$. The above function is zero when the message $m_2$ is not received. Now let us consider the case when message $m_1$ could not be received by receiver $Y_1$. This happens when $\mid H_2\mid > \mid H_1\mid$ or $\mid H_3\mid > \mid H_1\mid$, so we can derive, the function $\bar{\alpha}\left( .\right)$ is in the following form:
{\small
\begin{align}
\bar{\alpha}\left( K\right) \triangleq \left \{
\begin{matrix}
\bar{\alpha}^*\left( K\right)  & if: \mid H_1\mid >r \mid H_2\mid or  \mid H_1\mid > \mid H_3\mid \\ 0 &otherwise.
\end{matrix} \right.
\nonumber
\end{align}}
Here, again $\bar{\alpha}^*\left( .\right)$ is a deterministic function from $\mathcal{A}$ to $\left[ 0,1\right]$. As we said, for this function zero value occurs when the message $m_1$ is not received. However, here it is important to remind that our described channel is a less noisy channel which implies that for the channel to remain less noisy, we must have the following condition:
{\small
\begin{equation}
\label{H3<H2<H1}
\mid H_3 \mid \leq \mid H_2 \mid \leq \mid H_1 \mid . 
\end{equation}}

Now let us focus on the outer bound.

\subsection{Outer bound for the fading Gaussian 3-receiver less noisy broadcast channel}
Here, we use entropy power inequality (EPI) in a proper way to prove capacity outer bound for the fading Gaussian 3-receiver less noisy BC. Let $X$ and $Y$ be two independent complex random variables, according to EPI we have:
{\small
\begin{equation}
\label{EPI}
2^{H\left(X+Y\right)} \geq 2^{H\left( X\right)}+2^{H\left( Y\right)}, 
\end{equation}}  
also, we use the following equality for continuous complex random variable $X$:
{\small
\begin{equation}
\label{equality}
H\left( aX\right) =H\left( X\right) + \log \mid a\mid^2,
\end{equation}}
where, $a$ is a complex number.

\begin{theorem}
The outer bound for a fading Gaussian $3$-receiver less noisy broadcast channel $\left( \ref{fading-model} \right)$, is as below:
{\small
\begin{IEEEeqnarray}{lll}
\label{Converse-Gauss-fadin-less-noisy}
R_1\leq E\{ \psi \left[ \mid H_1\mid^2 \varphi \left( K\right) \bar{\vartheta}\left( \mathbf{S}\right) \right] \} \nonumber \\
R_2\leq E\{ \log \left[ \frac {\mid H_2\mid^2 \varphi \left( K\right) \bar{\gamma}\left( \mathbf{S}\right) +1}{\mid H_2\mid^2 \varphi \left( K\right) \bar{\vartheta}\left( \mathbf{S}\right) +1} \right] \} \nonumber \\
R_3\leq E\{ \log \left[ \frac {\mid H_3\mid^2 \varphi \left( K\right) +1}{\mid H_3\mid^2 \varphi \left( K\right) \bar{\gamma}\left( \mathbf{S}\right) +1} \right] \},
\end{IEEEeqnarray}}
for all arbitrary $\vartheta \left( . \right) $ and $\gamma \left( .\right)$ which are deterministic functions from $\mathcal{C}^3$ to $\left[ 0,1 \right]$ and for all power allocation functions $\varphi \left( .\right)$ from $\mathcal{A}$ to $R_+$ so that $E\{ \varphi \left( K\right) \}\leq P$.
\end{theorem}

\begin{IEEEproof}
First let us define the deterministic function $\varphi \left( .\right)$ as below:
{\small
\begin{equation}
\varphi \left( .\right) : \mathcal{A} \rightarrow R_+, ~~~ \varphi \left( k\right) \triangleq E\{ \mid X\mid^2 \mid K=k\}. \nonumber
\end{equation}}
Now to obtain $\left( \ref{Converse-Gauss-fadin-less-noisy}\right)$ we go through the similar procedure as explained in \cite{farsani}. To this end, for the first inequality from $\left( \ref{capacity-3-receiver-less-noisy-bc}\right)$ we have:
{\small
\begin{equation}
\label{10-maghale-khodam}
R_1\leq \int_\mathbf{s} I\left( X;Y_1\mid UV,\mathbf{s}\right)f\left( \mathbf{s}\right) d\mathbf{s},
\end{equation}}
where $I\left(  X;Y_1\mid UV,\mathbf{s}\right)$ can be written as:
{\small
 \begin{IEEEeqnarray}{lll}
 \label{11-maghale-khodam}
I\left(X;Y_1\mid UV,\mathbf{s}\right) =H\left( Y_1\mid UV,\mathbf{s}\right)-H\left(Y_1\mid UVX,\mathbf{s}\right) \nonumber \\
=H\left( Y_1\mid UV,\mathbf{s}\right)-H\left(h_1X+Z_1\mid UVX,\mathbf{s}\right) \nonumber \\
=H\left( Y_1\mid UV,\mathbf{s}\right)-\log \pi e,  
 \end{IEEEeqnarray}}
and we know according to the maximizing entropy property of Gaussian random variable for input power constraint we can bound $H\left( Y_1\mid UV, \mathbf{s}\right)$ as:
{\small
\begin{equation}
H\left( Y_1\mid UV,\mathbf{s}\right) \leq \log \pi e \left( \mid h_1\mid^2 \varphi \left( k\right) +1 \right), \nonumber
\end{equation}}
which can be changed into the following equality:
{\small
\begin{equation} 
\label{12-maghale-khodam}
H\left( Y_1\mid UV,\mathbf{s}\right)= \log \pi e \left( \mid h_1\mid^2 \varphi \left( k\right) \bar{\vartheta}\left( \mathbf{s}\right)+1\right), 
\end{equation}}
by considering a function $\bar{\vartheta}\left( .\right)$ as:
{\small
\begin{equation}
\bar{\vartheta}\left( .\right) : \mathcal{C}^3 \rightarrow \left[ 0,1\right]. \nonumber
\end{equation}}
So we can write $\left( \ref{11-maghale-khodam}\right)$ as:
{\small
\begin{equation}
\label{13-maghale-khodam}
I\left( X;Y_1 \mid UV,\mathbf{s}\right) =\log \left( \mid h_1\mid^2 \varphi \left( k\right) \bar{\vartheta}\left( \mathbf{s}\right) +1\right).
\end{equation}}
As a result, the first inequality in $\left( \ref{Converse-Gauss-fadin-less-noisy}\right)$ can be obtained by taking into account $\left( \ref{10-maghale-khodam}\right) -\left( \ref{13-maghale-khodam}\right)$.

Next for the second inequality from $\left( \ref{capacity-3-receiver-less-noisy-bc} \right)$ we have:
{\small
\begin{equation}
\label{14-maghal-khodam}
R_2\leq \int_\mathbf{s} I\left( V;Y_2\mid U,\mathbf{s}\right)f\left( \mathbf{s}\right) d\mathbf{s},
\end{equation}}
where $ I\left( V;Y_2\mid U,\mathbf{s}\right)$ can be written as:
{\small
\begin{equation}
\label{15-maghale-khodam}
 I\left( V;Y_2\mid U,\mathbf{s}\right) =H\left( Y_2\mid U,\mathbf{s}\right) -H\left( Y_2\mid UV,\mathbf{s}\right).
\end{equation}}
Also we know again from the maximizing entropy property of Gaussian random variable that:
{\small
\begin{equation}
H\left( Y_2\mid U,\mathbf{s}\right) \leq \log \pi e\left( \mid h_2\mid^2 \varphi \left( k\right) +1\right) , \nonumber
\end{equation}}
where the above inequality can be changed into the following equality:
{\small
\begin{equation}
\label{16-maghale-khodam}
H\left( Y_2\mid U,\mathbf{s}\right) =\log \pi e \left( \bar{\gamma}\left( \mathbf{s}\right) \mid h_2\mid^2 \varphi \left( k\right) +1\right),
\end{equation}}
by considering a function $\bar{\gamma} \left( . \right) :\mathcal{C}^3 \rightarrow \left[0,1 \right]$.

Before evaluating $H\left( Y_2\mid UV,\mathbf{s}\right)$, let $\tilde{Z}_1$ be a virtual noise which is a complex Gaussian random variable independent of noises $Z_1$, $Z_2$ and $Z_3$, with zero mean and unit variance. 
Now according to $\left( \ref{fading-model}\right)$ and $\left( \ref{H3<H2<H1}\right)$, $H\left( Y_2\mid UV,\mathbf{s}\right)$ can be written as:
{\small
\begin{equation}
H\left( Y_2\mid UV,\mathbf{s}\right) =H\left( \frac{h_2}{h_1}Y_1+ \sqrt{1-\mid \frac{h_2}{h_1}\mid^2}\tilde{Z}_1\mid UV,\mathbf{s}\right), \nonumber
\end{equation}}
on the other hand from EPI $\left( \ref{EPI}\right)$, $\left( \ref{equality}\right)$ and $\left( \ref{12-maghale-khodam}\right)$, we conclude:
{\small
\begin{IEEEeqnarray}{llll}
2^{H\left( Y_2\mid UV,\mathbf{s}\right)} =2^{H\left( \frac{h_2}{h_1}Y_1+ \sqrt{1-\mid \frac{h_2}{h_1}\mid^2}\tilde{Z}_1\mid UV,\mathbf{s}\right)} \nonumber \\
\geq 2^{H\left( \frac{h_2}{h_1}Y_1\mid UV,\mathbf{s}\right)}+2^{H\left( \sqrt{1-\mid \frac{h_2}{h_1}\mid^2}\tilde{Z}_1\mid UV,\mathbf{s}\right)} \nonumber \\
=\pi e \mid \frac{h_2}{h_1}\mid^2\left( \mid h_1\mid^2\varphi \left( k\right) \bar{\vartheta}\left( \mathbf{s}\right)+1\right) +\pi e\left( 1-\mid \frac{h_2}{h_1}\mid^2\right) \nonumber \\
=\pi e \left( \mid h_2\mid^2 \varphi \left( k\right) \bar{\vartheta}\left( \mathbf{s}\right)+1\right),\nonumber
\end{IEEEeqnarray}}
so from above inequality we have:
{\small
\begin{equation}
\label{19-maghale-khodam}
H\left( Y_2\mid UV,\mathbf{s}\right) \geq \log \pi e\left( \mid h_2\mid^2 \varphi \left( k\right) \bar{\vartheta}\left( \mathbf{s}\right) +1\right).
\end{equation}}
Now from $\left( \ref{15-maghale-khodam}\right)$, $\left( \ref{16-maghale-khodam}\right)$ and $\left( \ref{19-maghale-khodam} \right)$ it can be derived that:
{\small
\begin{equation}
\label{20-maghale-khodam}
I\left( V;Y_2\mid U,\mathbf{s}\right) \leq \log \left( \frac{\mid h_2\mid^2\varphi \left( k\right) \bar{\gamma}\left( \mathbf{s}\right) +1}{\mid h_2\mid^2\varphi \left( k\right)\bar{\vartheta}\left( \mathbf{s}\right) +1}\right).
\end{equation}}
As a result, the second inequality in $\left( \ref{Converse-Gauss-fadin-less-noisy} \right)$ can be obtained by considering $\left( \ref{14-maghal-khodam}\right)$ and $\left( \ref{20-maghale-khodam}\right)$.

For the third inequality in $\left( \ref{Converse-Gauss-fadin-less-noisy}\right)$ we go through the same steps as for the second one. From $\left( \ref{capacity-3-receiver-less-noisy-bc}\right)$ we have:
{\small
\begin{equation}
\label{21-maghale-khodam}
R_3\leq \int_\mathbf{s}I\left( U;Y_3\mid \mathbf{s}\right)f\left( \mathbf{s}\right) d\mathbf{s}.
\end{equation}}
We continue as follows:
{\small
\begin{equation}
\label{22-maghale-khodam}
I\left( U;Y_3\mid \mathbf{s}\right) =H\left( Y_3\mid \mathbf{s}\right) -H\left( Y_3\mid U,\mathbf{s}\right),
\end{equation}}
and it is obvious that:
{\small
\begin{equation}
\label{23-maghale-khodam}
H\left( Y_3\mid \mathbf{s}\right) \leq \log \pi e\left( \mid h_3\mid^2 \varphi \left( k\right) +1\right),
\end{equation}}
and for the second expression in $\left( \ref{22-maghale-khodam}\right)$, again we consider $\tilde{Z}_2$ as a complex Gaussian virtual noise, independent of noises $Z_1$, $Z_2$ and $Z_3$, with zero mean and unit variance. From $\left( \ref{fading-model}\right)$ and $\left( \ref{H3<H2<H1}\right)$ we can write:
{\small
\begin{equation}
\label{24-maghale-khodam}
H\left( Y_3\mid U,\mathbf{s}\right) =H\left( \frac{h_3}{h_2}Y_2+ \sqrt{1-\mid \frac{h_3}{h_2}\mid^2}\tilde{Z}_2\mid U,\mathbf{s}\right). \nonumber
\end{equation}}
So in a similar manner to the second inequality, using $\left( \ref{16-maghale-khodam} \right)$, EPI $\left( \ref{EPI}\right)$ and $\left( \ref{equality}\right)$ we can see:
{\small
\begin{equation}
\label{25-maghale-khodam}
H\left( Y_3\mid U, \mathbf{s}\right) \geq \log \pi e\left( \mid h_3\mid^2 \varphi \left( k\right) \bar{\gamma}\left( \mathbf{s}\right) +1\right),
\end{equation}}
so from $\left( \ref{22-maghale-khodam}\right)$, $\left( \ref{23-maghale-khodam}\right)$ and $\left( \ref{25-maghale-khodam}\right)$ we have:
{\small
\begin{equation}
\label{26-maghale-khodam}
I\left( U;Y_3\mid \mathbf{s}\right) \leq \log \left( \frac{\mid h_3\mid^2\varphi \left( k\right)+1}{\mid h_3\mid^2 \varphi \left( k\right) \bar{\gamma} \left( \mathbf{s}\right)+1}\right).
\end{equation}}
So from $\left( \ref{21-maghale-khodam} \right)$ and $\left( \ref{26-maghale-khodam}\right)$ we derive the third inequality in $\left( \ref{Converse-Gauss-fadin-less-noisy} \right)$.
\end{IEEEproof}


\subsection{Capacity of fading Gaussian $3$-receiver less noisy broadcast channel under certain conditions}
For the following case the regions $\left( \ref{achiev-Gaussian-fadin-3-less}\right)$ and $\left( \ref{Converse-Gauss-fadin-less-noisy}\right)$ coincide.

\begin{theorem}
When we have the perfect state information at the transmitter i.e. $K\equiv \mathbf{S}$, the state of the channel is known to the transmitter completely. In this special case, the inner and outer bounds of fading Gaussian 3-receiver less noisy broadcast channel coincide and we will have the following capacity region:
{\small
\begin{IEEEeqnarray}{lll}
\label{capacity-special-case-Gaussian-fad}
R_1\leq E\{ \psi \left[ \mid H_1\mid^2 \varphi \left( \mathbf{S}\right) \bar{\alpha}\left( \mathbf{S}\right) \right] \} \nonumber \\
R_2\leq E\{ \psi \left[ \frac{\mid H_2 \mid^2 \varphi \left(\mathbf{S}\right) \alpha \left(\mathbf{S}\right) \bar{\beta}\left( \mathbf{S}\right)}{\mid H_2\mid^2 \bar{\alpha}\left( \mathbf{S}\right) \varphi \left( \mathbf{S}\right) +1} \right] \} \nonumber \\
R_3\leq E\{ -\psi \left[ \frac{-\mid H_3\mid^2 \varphi \left( \mathbf{S}\right) \alpha \left( \mathbf{S}\right) \beta \left( \mathbf{S}\right)}{\mid H_3\mid^2 \varphi \left( \mathbf{S}\right) +1} \right]\},
\end{IEEEeqnarray}}
for all arbitrary $\alpha \left( .\right)$ and $\beta \left( .\right)$ which are deterministic functions from $\mathcal{C}^3$ to $\left[0,1 \right]$ and have a specific relation with each other and for all power allocation functions $\varphi \left( .\right)$ from $\mathcal{C}^3$ to $R_+$ so that $E\{ \varphi \left( \mathbf{S}\right) \}\leq P$.
\end{theorem}

\begin{IEEEproof}
Here we have perfect CSIT, hence $\alpha \left( . \right)$ and $\beta \left( . \right)$ are two deterministic functions from $\mathcal{C}^3$ to $\left[0,1\right]$. Now let us compare corresponding inequalities in $\left( \ref{achiev-Gaussian-fadin-3-less}\right)$ and $\left( \ref{Converse-Gauss-fadin-less-noisy} \right)$. As we see the first inequalities in $\left( \ref{achiev-Gaussian-fadin-3-less}\right)$ and $\left( \ref{Converse-Gauss-fadin-less-noisy} \right)$ will be the same when we put $K\equiv \mathbf{S}$ and $\alpha \left( \mathbf{s}\right) =\vartheta \left( \mathbf{s}\right)$. But for second inequalities let us return to $\left( \ref{Converse-Gauss-fadin-less-noisy} \right)$. As we see if we have the function $\bar{\gamma}\left( .\right)$ in $\left( \ref{16-maghale-khodam}\right)$ as below:
{\small
\begin{equation}
\bar{\gamma}\left( \mathbf{s}\right)=1-\alpha \left( \mathbf{s}\right) \beta \left( \mathbf{s}\right),
\nonumber
\end{equation}}
then, according to the fact that $1-\alpha \left(  \mathbf{s}\right) \beta \left(  \mathbf{s}\right) =\bar{\alpha}\left(  \mathbf{s}\right) +\alpha \left(  \mathbf{s}\right) \bar{\beta}\left(  \mathbf{s}\right)$ the second inequality in $\left( \ref{Converse-Gauss-fadin-less-noisy}\right)$ changes into the form of its counterpart in $\left( \ref{achiev-Gaussian-fadin-3-less}\right)$. By considering above equality for deterministic function $\bar{\gamma}\left( .\right)$, it is easily seen that the third inequality in $\left( \ref{Converse-Gauss-fadin-less-noisy}\right)$ matches its counterpart in $\left( \ref{achiev-Gaussian-fadin-3-less}\right)$.
\end{IEEEproof}

\section{Conclusion}
In this paper, we investigated special classes of both discrete and continuous alphabet $3$-receiver broadcast channels. In presence of non-causal side information, first we derived capacity bounds for discrete memoryless multilevel broadcast channel then, we obtained capacity region for discrete memoryless $3$-receiver less noisy broadcast channel. Also we showed that, for this discrete alphabet cases, our obtained regions reduce to previous results of some studied channels. Finally, we obtained capacity bounds for the fading Gaussian $3$-receiver less noisy broadcast channel with partial channel state information at the transmitter (CSIT), introducing an efficient signaling for achievability proof and using EPI in a proper way  for converse proof. Furthermore, we showed that for the special case in which we have perfect CSIT for this fading Gaussian channel these bounds coincide under certain conditions and result in the capacity. 
\bibliographystyle{IEEEtran}
\bibliography{g1}

\end{document}